\documentclass[useAMS,usenatbib]{mn2e}
\usepackage{graphicx}

\newcommand{\be}{\begin{equation}}
\newcommand{\ee}{\end{equation}}
\newcommand{\bea}{\begin{eqnarray}}
\newcommand{\eea}{\end{eqnarray}}

\title[NoSOCS in SDSS. II]{NoSOCS in SDSS. II. Mass Calibration of Low Redshift Galaxy Clusters with Optical and X-ray Properties}
\author[Lopes et al.]{P. A. A. Lopes$^{1}$\thanks{E-mail: 
paal@univap.br}, R. R. de Carvalho$^{2}$, J. L. Kohl-Moreira$^{3}$,
C. Jones$^{4}$\\
$^{1}$IP\&D Universidade do Vale do Para{\' \i}ba, 
Av. Shishima Hifumi 2911, \\ 
S\~ao Jos\'e dos Campos, SP 12244-000, Brasil \\ 
$^{2}$Instituto Nacional de Pesquisas Espaciais -- Divis\~ao de
Astrof{\' \i}sica (CEA), Avenida dos Astronautas, 1758\\
S\~ao Jos\'e dos Campos, SP 12227-010, Brasil \\
$^{3}$Observat\'orio Nacional/MCT, COAA, Brasil \\
$^{4}$Harvard-Smithsonian Center for
Astrophysics, 60 Garden Street, Cambridge, MA 02138, USA}
\begin{document}

\date{Accepted  Received ; in original form }

\pagerange{\pageref{firstpage}--\pageref{lastpage}} \pubyear{2009}

\maketitle

\label{firstpage}

\begin{abstract}
We use SDSS data to investigate the scaling relations of 127 NoSOCS
and 56 CIRS galaxy clusters at low redshift ($z \le 0.10$).  We show
that richness and both optical and X-ray luminosities are reliable
mass proxies. The scatter in mass at fixed observable is $\sim$ 40\%,
depending on the aperture, sample and observable considered. For 
example, for the
massive CIRS systems $\sigma_{lnM500|N500}$ = 0.33 $\pm$ 0.05 and
$\sigma_{lnM500|Lx}$ = 0.48 $\pm$ 0.06. For the full sample 
$\sigma_{lnM500|N500}$ = 0.43 $\pm$ 0.03 and 
$\sigma_{lnM500|Lx}$ = 0.56 $\pm$ 0.06. The scaling relations based only 
on the richer systems (CIRS) are slightly flatter than those based on the 
full sample, but the discrepancies are within 1-$\sigma$. We estimate 
substructure using two and three dimensional optical data, verifying that
substructure has no significant effect on the cluster scaling
relations (intercepts and slopes), independent of which
substructure test we use. For a subset of twenty-one clusters, we
estimate masses from the M-T$_X$ relation using temperature measures
from BAX. The scaling relations derived from the optical and X-ray
masses are indeed very similar, indicating that our method
consistently estimates the cluster mass and yields equivalent results
regardless of the wavelength from which we measure mass. For massive systems, 
we represent the mass-richness relation by a function with the form 
${\rm ln (M_{200}) = A + B \times ln(N_{200}/60)}$,
with M$_{200}$ being expressed in units of 10$^{14}$ M$_{\odot}$. Using
the virial mass, for CIRS clusters, we find A = (1.39 $\pm$ 0.07) and
B = (1.00 $\pm$ 0.11). For the same sample, but using the masses obtained 
by the caustic method, we get A = (0.64 $\pm$ 0.14) and B = (1.35 $\pm$ 0.34).
If we consider the mass as estimated from T$_X$ (for the subset of 21 clusters
with T$_X$ available) we derive A = (0.90 $\pm$ 0.10) and B = (0.92 $\pm$ 0.10).
The relations based on the virial mass have a scatter of $\sigma_{lnM200|N200}$ 
 = 0.37 $\pm$ 0.05, while $\sigma_{lnM200|N200}$ = 0.77 $\pm$ 0.22 for
the caustic mass and $\sigma_{lnM200|N200}$ = 0.34 $\pm$ 0.08 for the temperature
based mass.
\end{abstract}


\section{Introduction}

Two aspects of galaxy clusters make their study uniquely important for
cosmology. The path from density perturbations to virialized objects
involves the physical mechanisms that ultimately determine the cluster
properties as we currently observe them. The growth of these systems
depends on cosmological parameters like $\Omega_{\rm m}$,
$\Omega_{\lambda}$, $\sigma_{\rm 8}$, etc. The sensitivity to these
parameters depends on the redshift \citep{car92}. 
Therefore, the systematic study of cluster properties,
based on unbiased samples at different redshifts, can provide
substantial input for models trying to explain not only how
clusters formed and evolved from the astrophysical viewpoint but also
how spacetime is structured (e.g. \citealt{hen91, ros02}). 
Ultimately, we aim to disentangle the astrophysical aspects 
from the cosmological ones.

Clusters have no sharp edges. Observations show that their central parts 
reached virial equilibrium earlier while the outskirts are still accreting 
matter from their surroundings, making the measurements 
of cluster properties a 
difficult issue.  Clusters are constituted of both dark (non-baryonic) and 
luminous (baryonic) matter. This mixture renders observations within a 
specific wavelength range very limited in utility and determining 
the physical state of each component too complex. 
Over the years, the sizes of cluster samples and the
quality of the data hampered our progress on all of the fronts mentioned 
above. Only recently have large samples with high-quality data been 
obtained, allowing a systematic investigation of these effects 
(e.g. \citealt{vik08, hen04}).

The abundance of clusters of galaxies can be used as a probe of linear 
density fluctuations \citep{eke96} and since their growth is directly 
determined from the dark energy properties, measuring the number of clusters 
of a given mass as a function of redshift can constrain the equation 
of state (${\sl w}$ parameter) of this majority component of the Universe 
(e.g. \citealt{fri08}). As shown by \citet{hut01}, the 
sensitivity of cluster abundance to ${\sl w}$ is 
maximum around redshift 1.0, so that a cluster sample over a large 
area on the sky and deep (z $\sim$ 1.0) is critically needed to provide 
reliable results. Direct measures of halo mass are not possible, so that
we have to rely on mass proxies free of systematics. The most accurate
mass tracers are velocity dispersion and X-ray temperature. Unfortunately,
for large samples, at any redshift, directly measuring these cluster 
properties is not feasible. Hence, we have to rely on other observables
(such as richness or X-ray luminosity) that are easier to derive for
all clusters in a given survey.

Several studies have investigated mass-observable (MO) relations and 
their evolution with redshift in the optical and X-ray 
\citep{vik02, lev02, ett04, sta06, bec07, ryk08}. As advocated 
by \citet{lim04, lim05}
the variance of the counts themselves (clustering 
of clusters)  can be used to calibrate the MO relation. Dark energy 
depends only on the redshift, but the cluster properties vary with mass 
and redshift. Using cosmological simulations one can use counts and their 
variance to not only normalize the MO relation but also learn which 
cosmology fits the data better. Therefore, understanding the scaling 
relations of galaxy clusters and their dependence on the dynamical 
state of the system is crucial. 

This paper focuses on four specific issues: (i) examining the
performance of different cluster properties, like richness (N$_{\rm gals}$),
optical (L$_{\rm opt}$) and X-ray (L$_{\rm X}$) luminosities\footnote{The error in
the X-ray luminosity presented in the first paper of this series was 
underestimated. That is fixed in the current paper. All the X-ray 
measurements available in Table 5 of the first paper are replaced by the 
values listed in Table A1 of this paper.}, 
as proxies for cluster mass. The
relation between one of these proxies and an independent mass estimate
defines a scaling relation; (ii) measuring how much substructure there
is in clusters (in 2D and 3D) and establishing its effect on the final
mass calibration; (iii) investigating how the scaling relations change
when mass is derived from the analysis of the velocity distribution or
temperature. Optical masses used in this work are either from a virial
analysis (ours) or from the caustic technique (\citealt{rin06},
hereafter RD06); and (iv) comparing the optical and X-ray properties of
galaxy clusters, as well as obtaining a direct relation between R$_{200}$
and richness.

This paper is organized as follows: $\S$2 briefly describes 
the samples used here, while $\S$3 presents the mass calibration based on
N$_{\rm gals}$, L$_{\rm opt}$ and L$_{\rm X}$. In this section we also investigate
the effects of substructure on the scaling relations and compare the
results based on different mass estimators. These are derived with optical 
data, from the virial or caustic analysis, or from the gas temperature. The
mass-to-light ratio is also presented at the end of this section.
In $\S$4 we discuss how our results compare to others in the literature. 
Correlations
between optical and X-ray properties are presented in $\S$5, while in $\S$6
we establish the connection between richness and physical radius. The main 
conclusions are drawn in $\S$7. The cosmology assumed in this work is 
$\Omega_{\rm m}=$0.3,  $\Omega_{\lambda}=$0.7, and 
H$_0 = 100$ $\rm h$ $\rm km$ $s^{-1}$ Mpc$^{-1}$, with $\rm h$ set to 0.7.

\section{Data and Methods}
The main data set used in this work is the supplemental version of the
Northern Sky Optical Cluster Survey (NoSOCS, \citealt{lop03, lop04}),
which has its origin on the digitized version of the Second Palomar
Observatory Sky Survey (POSS-II; DPOSS, \citealt{djo03}). In
\citet{gal04} and \citet{ode04}, we describe photometric calibration
and object classification for DPOSS, respectively. The supplemental
version of NoSOCS \citep{lop04} goes deeper ($z \sim 0.5$), but covers
a smaller region than the main NoSOCS catalog \citep{gal03, gal09}, and
contains 9,956 cluster candidates over $\sim$ 2,700 square
degrees. The smaller area is due to the use of the best DPOSS plates,
selected according to seeing and limiting magnitude ($r = 21.0$). 

We examine a sample of  low redshift galaxy clusters ($z \le 0.10$) 
from the NoSOCS supplemental catalog. As this survey 
comprises mainly poor systems in this redshift
range, we complemented them with more massive systems
from the Cluster Infall Regions in SDSS (CIRS) sample
(RD06). CIRS is a collection of $z \le 0.10$ X-ray selected clusters
overlapping the SDSS DR4 footprint. As described in the first 
paper of this series (\citealt{lop09}) we extracted SDSS data for 127 NoSOCS 
clusters and 56 CIRS systems at low-$z$. 

We used photometric and spectroscopic data from the fifth release
(DR5) of the Sloan Digital Sky Survey \citep{yor00}. The exception is
for the CIRS systems, which were incorporated in a late stage of this
work,  thus having data from the SDSS DR6. All the magnitudes retrieved
from SDSS are de-reddened model magnitudes. Details regarding SDSS
data extraction are provided in paper I \citep{lop09}. There, we used
SDSS photometric data to estimate more accurate photometric redshifts
($z_{photo}$, \citealt{lop07}), richnesses and optical luminosities for the 
full NoSOCS supplemental catalog. We found 7,414
systems well sampled in SDSS DR5. Approximately 10\% (754) have
$z_{photo} \le$ 0.133 \citep{lop09}.

The above redshift limit comes from the choice of only using 
clusters at $z \le 0.10$
for the current work. In this redshift range the SDSS spectroscopic survey 
is complete. As discussed in \citet{lop09},
at higher redshifts galaxies fainter than $M^*+1$ are missed, biasing the 
dynamical analysis (see discussion in section 4.3 of \citealt{lop09}).
Out of the 754 NoSOCS supplemental clusters with $z_{photo} \le$ 0.133, 
we were able to determine the spectroscopic redshift
for 179 systems, requiring at least 3 galaxies within 0.50 h$^{-1}$ Mpc in 
the SDSS spectroscopic footprint. Note that we only selected systems for 
which $|z_{photo} - z_{spec}| \le 0.03(1+z_{spec})$. 
We eliminated interlopers using 
the ``shifting gapper'' technique \citep{fad96}, applied to all 
galaxies with spectra available within 2.50 h$^{-1}$ Mpc. From the 
179 clusters with $z_{spec} \le 0.10$  
we retained 127 systems with at least 10 member galaxies selected 
by the above procedure. We applied the same procedure to
the 56 CIRS clusters.
Figures 4 and 5 of \citet{lop09} show the velocity-radius distributions of
the 127 NoSOCS clusters and the 56 CIRS systems.

These clusters were then subjected to a virial analysis analogous to
the one described in \citet{gir98}, \citet{pop05, pop07} and
\citet{biv06}. This procedure yields estimates of $\sigma_P$,
$R_{500}$, $R_{200}$, $M_{500}$ and $M_{200}$ for the set of 183 low
redshift clusters considered in this work. X-ray luminosity is also
estimated for these systems using ROSAT All Sky Survey (RASS) data. A
detailed description of both the virial analysis and derivation of
X-ray luminosity is provided in \citet{lop09}, which also lists the
properties of these systems in Tables 1, 2, 3, 4, and 5.

The NoSOCS clusters have velocity dispersion estimates of $100 <
\sigma_P < 700$ km/s. The CIRS systems have $200 < \sigma_P < 900$ km/s
(with only 23\% of objects with $\sigma_P < 400$ km/s). For most of the
studies below we ignore three of the CIRS objects which have biased
values of $\sigma_P$ and mass due to projection effects
and substructure (see paper I). We only show these objects in 
Figures 9, 10, and
11, where we investigate the impact of substructure on the scaling
relations. These 3 clusters are Abell 1035B, Abell 1291A and Abell
1291B.
\section{Mass Calibration}

Scaling relations involving a simple cluster observable, like richness
or luminosity and a fundamental property such as mass, can provide
important clues on how large scale structure forms, galaxy formation
proceeds, and how the intra-cluster gas reaches its current state. The
establishment of these relations for nearby clusters is also vital for
future studies of the cluster population in the distant
universe. Particularly, they may help constrain the dark energy
equation of state \citep {maj04}.

In this section, we examine the correlation of N$_{gals}$, L$_{opt}$ and 
L$_X$ with cluster properties, such as velocity dispersion ($\sigma_P$) and 
mass (M$_{200}$ and M$_{500}$). We use the spectroscopic 
redshifts of the clusters (instead of $z_{photo}$) when computing richnesses 
and luminosities. Three different apertures are used to derive 
these quantities: 0.50 h$^{-1}$ Mpc, R$_{500}$ and R$_{200}$. We consider 
the re-centered (luminosity-weighted) coordinates for the 127 NoSOCS clusters 
and the original position (X-ray centroid) listed in RD06 for the 
56 CIRS clusters.  We present the results for the NoSOCS and CIRS samples 
independently as well as for the combined sample. 

In the next three subsections we discuss the mass calibration based
independently on richness, optical luminosity, and X-ray luminosity.
For the richness relations we also investigate the impact of using
less rigorous criteria for interloper removal when computing mass, as
well as assuming different centroids and considering all NoSOCS
clusters at $z \le 0.25$. We then examine the impact of cluster substructure
on the scaling relations and compare the results derived from
mass estimates obtained from different wavelengths. Finally, we show the
dependence of the mass-to-luminosity ratio with the cluster scale (defined
by its mass).

\subsection{Calibration with richness}

Figure 1 shows the comparison between mass (M$_{500}$ and M$_{200}$) with
richness, estimated within R$_{500}$ and R$_{200}$. The lower
panels show the residuals, LOG(M$_{obs}$/M$_{fit}$). M$_{obs}$ 
is the observed value
of the mass within the given radius, while M$_{fit}$ is 
the linear regression solution.
The results are only shown for the 127 NoSOCS objects. 
The solid line in each upper panel shows the orthogonal regression fit 
\citep {akr96}. All the scaling relations obtained in this work are of the form
\bea
{\rm ln (Y)} = {{\rm A + B \times ln({{X \over C}})}} \,\,\, ,
\eea
The X and Y parameters are listed in the tables defining the relations (see 
below). The pivot point (C) depends on the sample being used. It is taken to 
be approximately the median value of the X parameter 
within R$_{500}$ for a given sample. For the NoSOCS and full samples (NoSOCS 
+ CIRS) we consider the same pivot point, while for the more massive 
systems (CIRS) we have another pivot point. Those are listed in the tables
and can be seen in the figures. The linear fit is obtained by a two iteration 
3-sigma clipping, so that after a first run we eliminate outliers. Those are 
not used in the final fit and are indicated (if exist) as open symbols in the
figures.

\begin{figure*}
\begin{center}
\leavevmode
\includegraphics[width=6.0in]{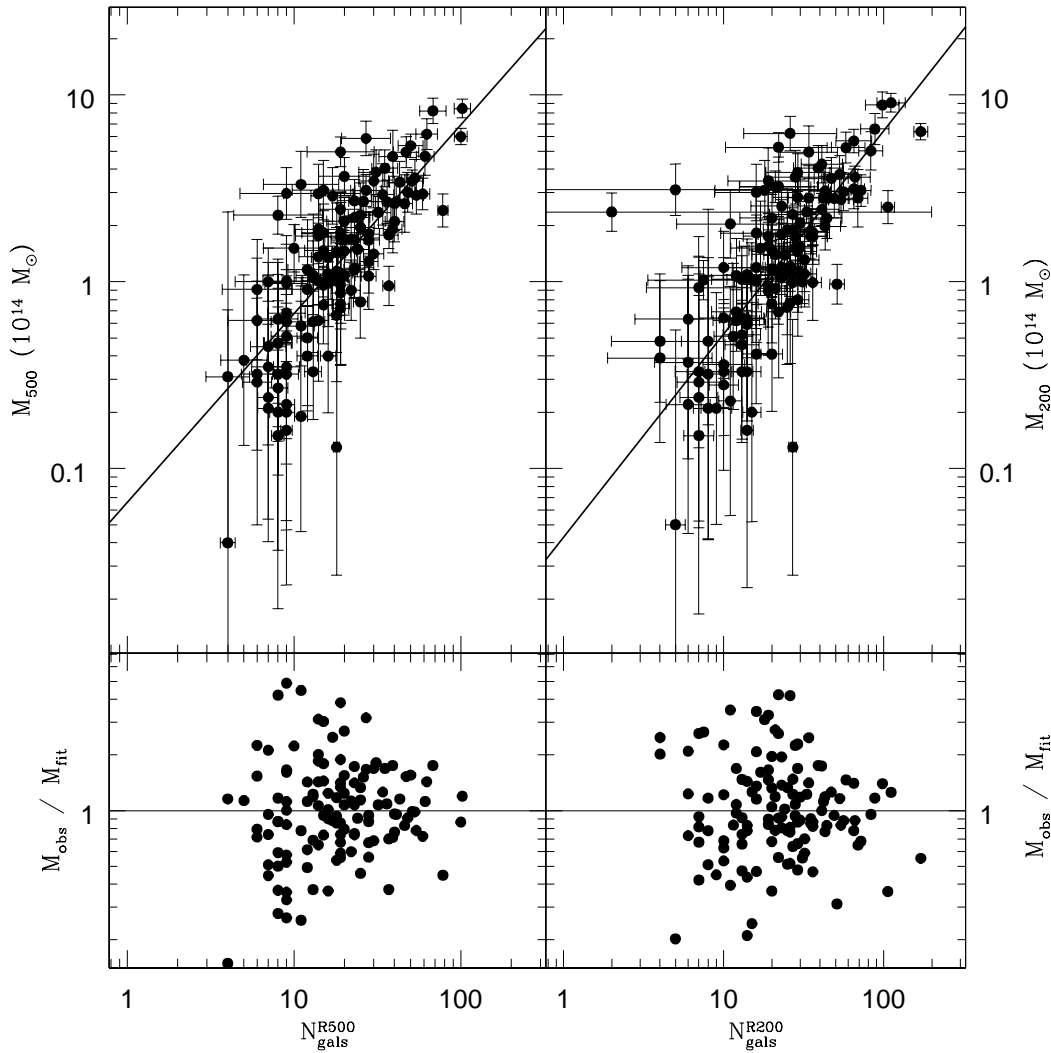}
\end{center}
\caption{The connection between mass and richness, computed with two
apertures (R$_{500}$ and R$_{200}$). In the lower panels the residuals
LOG(M$_{obs}$/M$_{fit}$) are shown. M$_{obs}$ is the observed value
of the mass, while M$_{fit}$ is the linear regression result.}
\label{fig:richscalerelations}
\end{figure*}

The results for the scaling relations involving $\sigma_P$, M$_{500}$,
M$_{200}$ and N$_{gals}$ are summarized in Table 1. The columns give
(1) the parameters involved in the relation (the abscissa and ordinate
of the corresponding plots); (2) the cluster sample used; (3) and (4)
the intercept (A) and slope (B); (5) the scatter (in natural log space) in 
the Y parameter at fixed X; (6) the total number of
clusters used in the preliminary fit; and (7) the final number after a
3-$\sigma$ clip. Note that we evaluated the error in the scatter, performing
a boostrap procedure (500 events) with replacement. So the scatter is shown
with the associated error in all tables (exception to Table 8).
Rows 1-6 provide the results using only the NoSOCS
clusters, as shown in Figure 1. Rows 7-12 are for clusters from RD06
only,  while the combination of NoSOCS and RD06 is listed in rows
13-18.

If one has to work within a fixed metric radius, 0.50 h$^{-1}$ Mpc
yields the smallest error in the richness or L$_{opt}$ measures
and minimizes the scatter in the optical versus X-ray scaling
relations \citep{pop04, lop06}. However, when compared to apertures
that scale with mass, this radius (0.50 h$^{-1}$ Mpc) is too
large for low mass systems and too small for massive ones, and may lead
to tilted (steeper) scaling relations (see Table 1). When we use an
aperture that scales with mass the relations become
flatter. For instance, for the NoSOCS sample, the slope of the 
M$_{200}$-N$_{gals}$ relation is 1.45 when using the fixed metric radius 
for computing richness (0.50 h$^{-1}$ Mpc), but only 1.09 when using R$_{200}$.

From the lower panels of Figure 1 we see that some of the poorest systems
(N$_{gals}^{R200} < 30$) have larger
deviations from the linear solution. Both richness and mass are harder 
to determine due to the low contrast of some of these clusters.
Incompleteness in spectroscopic sampling for these poor systems may
also be a problem (see discussion in $\S$3.2).

The comparison of the relations for the 127 NoSOCS clusters, the
53 CIRS systems and the combined sample, shows that the scatter
of the relations obtained with the CIRS and extended samples is
smaller than for the NoSOCS clusters alone. This is because CIRS is
restricted to more massive systems and the enlarged sample has better
statistics, spanning a wider range of all parameters. For the
M$_{200}$-N$_{gals}^{R200}$ relation the orthogonal scatter in mass at
fixed richness is (0.60 $\pm$ 0.07) for NoSOCS, (0.37 $\pm$ 0.05) for the 
CIRS sample and (0.46 $\pm$ 0.04) for the combined data set. 
The slopes are nearly the same for the NoSOCS
and extended samples, but slightly different for CIRS, although they
are compatible within the errors. For the M$_{200}$-N$_{gals}^{R200}$
relation the slope is 1.09 for the NoSOCS sample, 1.00 for CIRS
and 1.07 for the combined sample.

\subsubsection{Systematics in the scaling relations}

Table 2 lists the coefficients of the relations
$\sigma_P$-N$_{gals}^{R200}$ and M$_{200}$-N$_{gals}^{R200}$ using
only the NoSOCS clusters, with three different methods. In the first
two rows we consider the original cluster coordinates instead of the
luminosity-weighted ones. Rows 3-4 list the results when we use a more
``relaxed'' criteria for interloper removal (see $\S$ 4.1 of \citealt
{lop09}), and in rows 5-6 we consider all NoSOCS clusters at $z \le$
0.25, instead of $z \le$ 0.10 (limit adopted for this work). The
latter is useful to see how much the scaling relations are affected
for using clusters at redshifts where the spectroscopic sampling is
very incomplete (see $\S$ 4.3 of paper I). Note that \citet{pop05}
(hereafter POP05) considered clusters at $z \le 0.25$ for deriving
scaling relations based on SDSS data.

When comparing the results obtained with the original coordinates
(first two rows of Table 2) to the ones with the
luminosity-weighted centroids (Table 1), we find that the intercept,
slope and scatter are consistent within the errors (1-$\sigma$). Thus, we 
conclude that the centroid is not a critical issue when deriving the scaling
relations. When the criteria to select interlopers are relaxed
(results in rows 3-4 of Table 2) the agreement is not as good for the
intercept and slope, but it is still within 1.5-$\sigma$. Relaxing the 
criteria for rejecting interlopers in the ``shifting-gapper'' technique has 
a minor effect in the relations. 

The third case in Table 2 uses all NoSOCS clusters at $z \le$ 0.25
(rows 5-6). A comparison to Table 1 reveals that the intercept and
slope are now different, in some cases consistent only within
2-$\sigma$. The scatter is, however, similar. This result shows the
relevance of using complete spectroscopic samples for studying the
scaling relations. The use of clusters at redshifts where the
spectroscopic survey is not complete to M$^* + 1$ results in biased
velocity dispersion and mass estimates (see paper I), yielding flatter
scaling relations.

This is a critical issue for measuring the cluster mass function since
the mass calibration might be severely biased. However, it is
important to note that this bias in the calibration may reflect the
percentage of clusters with incomplete spectroscopic sampling. In this
work 127 of the 219 NoSOCS clusters at $z \le$ 0.25 are below $z \le$
0.10, and are therefore well sampled. Ninety-two of the 219 systems
(42\%) have poor sampling ($z >$ 0.10), a considerable fraction of the
total used when deriving the scaling relations. Other authors included
$z >$ 0.10 clusters in their sample (POP05, for instance), but show
results consistent to our unbiased sample (at $z \le$ 0.10). Although
they initially have many systems at $z >$ 0.10, they require that clusters have
at least ten galaxies with redshifts when estimating mass and deriving
the scaling relations, which may preferentially exclude many of the
higher redshift clusters. We believe that this is the case since their
results are consistent with ours.

\subsection{Calibration with optical luminosity}

\begin{figure*}
\begin{center}
\leavevmode
\includegraphics[width=6.0in]{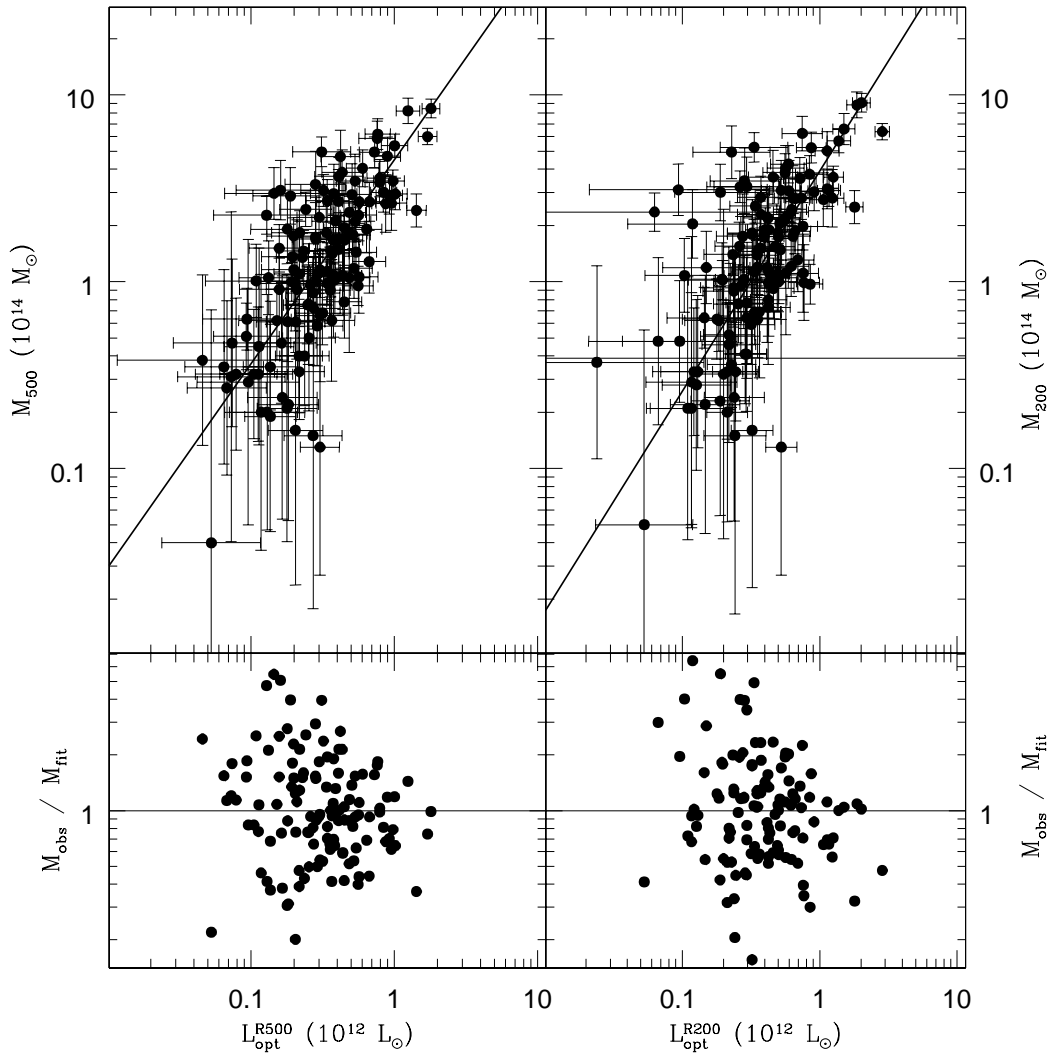}
\end{center}
\caption{Analogous to Figure 1, but for mass and optical luminosity.}
\label{fig:richscalerelations}
\end{figure*}

\begin{figure*}
\begin{center}
\leavevmode
\includegraphics[width=6.0in]{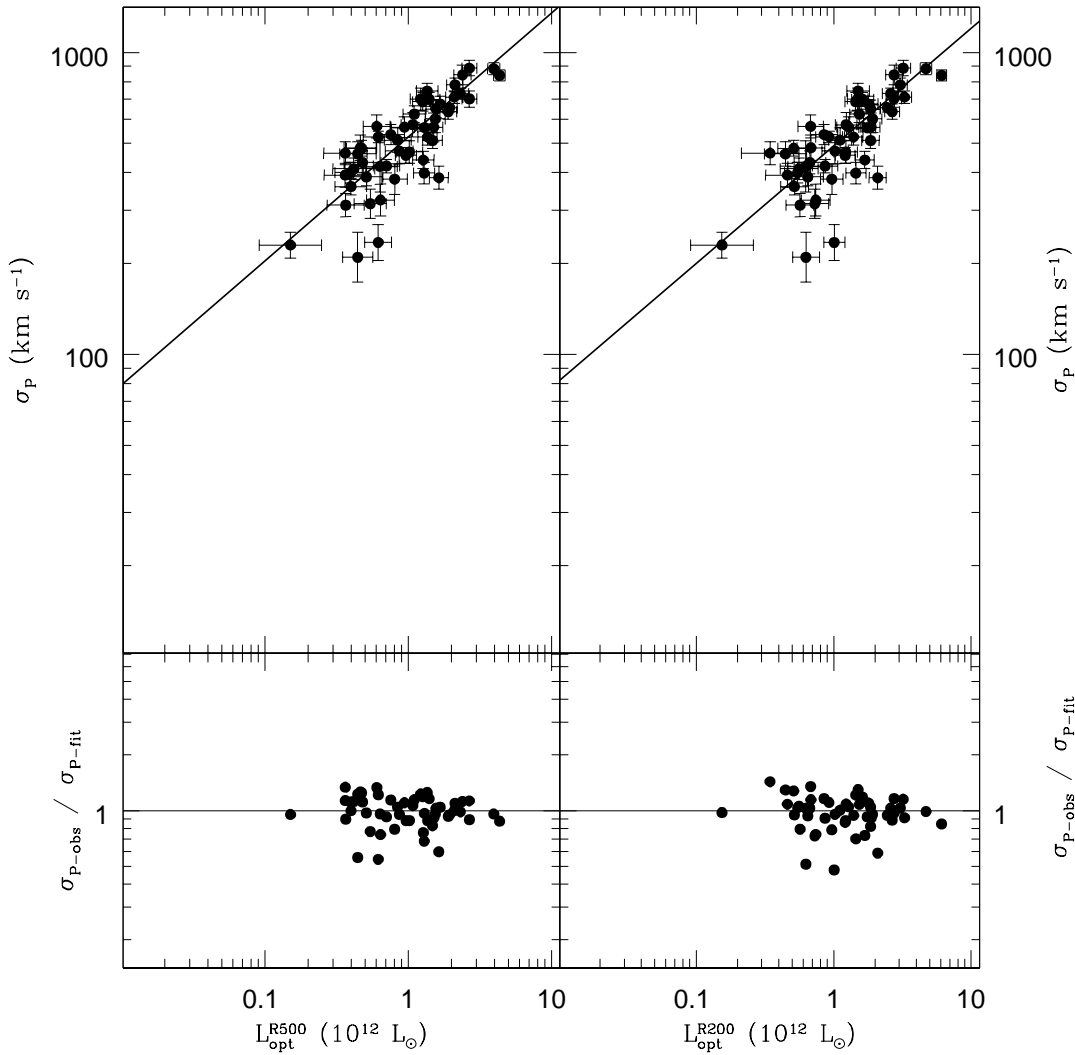}
\end{center}
\caption{Analogous to Figure 2, but using only the 53 clusters from 
RD06 and considering $\sigma_P$ and L$_{opt}$.}
\label{fig:richscalerelations}
\end{figure*}

\begin{figure*}
\begin{center}
\leavevmode
\includegraphics[width=6.0in]{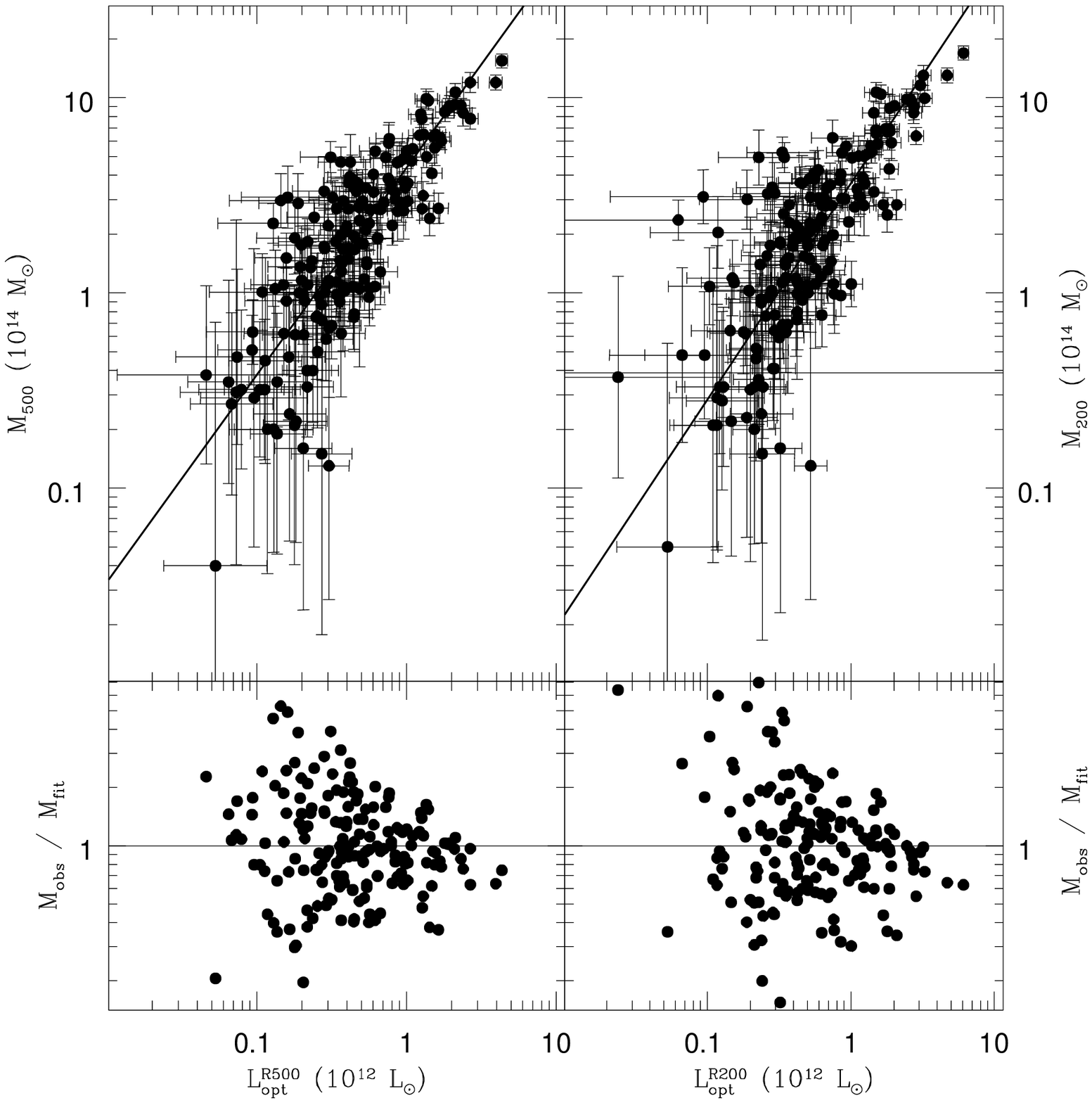}
\end{center}
\caption{Analogous to Figure 2, but for the 
combined sample of 127 NoSOCS clusters and 53 systems from RD06.} 
\label{fig:richscalerelations}
\end{figure*}

Figure 2 shows the comparison between mass (M$_{500}$ and M$_{200}$)
and optical luminosity computed within R$_{500}$ or
R$_{200}$. Residuals are shown in the lower panels. We only show the
results for the 127 NoSOCS objects. One interesting feature from the
comparison of Figures 1 and 2 (see Tables 1 and 3) is that the 
slopes obtained in the two
cases are consistent. This indicates the optical luminosity and the
number of galaxies used to compute L$_{\rm opt}$, are proportional to
each other (see \citealt {pop07}), with a constant of proportionality
consistent with unity. Figure 3 shows $\sigma_P$ versus L$_{opt}$,
computed within R$_{500}$ or R$_{200}$ for only the CIRS
objects. Figure 4 is analogous to Figures 2 and 3, but shows the extended
sample of 180 NoSOCS plus CIRS clusters. The inspection of the lower
panels of Figures 2 and  4 reveals that larger deviations from the linear
regression occur for L$_{opt}^{R200} < 0.30$ $10^{12}$ L$\odot$,
corresponding to the cut at N$_{gals}^{R200} = 30$ seen in Figure 1. Hence,
we stress the conclusion that richness, L$_{\rm opt}$ and mass are harder
to obtain due to the low contrast of such systems. Another problem affecting
these clusters may be incompleteness in the spectroscopic sampling. 

To confirm these conclusions we separate the 
clusters shown in Figure 4 in systems with
$|{\rm LOG(M_{obs}/M_{fit})}| \le 2$ or $|{\rm LOG(M_{obs}/M_{fit})}| > 2$. 
We find 135 clusters in
the first case and only forty-five (25\%) in the second. 
For $|{\rm LOG(M_{obs}/M_{fit})}| \le 2$ we
find the following median values: N$_{gals}$ = 33.0, 
N$_{gals-err}$/N$_{gals}$ = 0.24, N$_{spec}$ = 48.0 and N$_{spec200}$ = 21.0; 
where N$_{gals}$ is computed within R$_{200}$, N$_{gals-err}$/N$_{gals}$ is 
the percentage error on the richness measurement, N$_{spec}$ is the number of 
members (spectroscopically selected) within the maximum aperture (normally 
2.5 h$^{-1}$ Mpc; see paper I) 
and N$_{spec200}$ is the number of members within R$_{200}$. 
For $|{\rm LOG(M_{obs}/M_{fit})}| > 2$ we have N$_{gals}$ = 22.0, 
N$_{gals-err}$/N$_{gals}$ = 0.36, N$_{spec}$ = 30.0 and N$_{spec200}$ = 15.0
Thus, the forty-five clusters with larger mass deviations are generally poor, 
low contrast (indicated by their large percentage error in richness) and
have their mass estimates based on few galaxies. 

The results shown in Figures 2, 3 and 4 are summarized in Table 3. The
meaning of all columns and rows is analogous to those in Table
1. The same conclusions drawn for Table 1 are valid now. The most
important result is that the relations involving
L$_{opt}$ have approximately the same scatter as those obtained with
N$_{gals}$. In other words, optical luminosity performs as good as 
richness for mass calibration. For the full sample, the scatter of 
M$_{200}$ at fixed N$_{gals}^{R200}$ is (0.46 $\pm$ 0.04), while it is 
(0.49 $\pm$ 0.04) at fixed L$_{opt}^{R200}$. For the CIRS sample, the
scatter of M$_{200}$ at fixed N$_{gals}^{R200}$ is (0.37 $\pm$ 0.05), being
(0.36 $\pm$ 0.06) at fixed L$_{opt}^{R200}$.

\subsection{Calibration with X-ray luminosity}

\begin{figure*}
\begin{center}
\leavevmode
\includegraphics[width=6.0in]{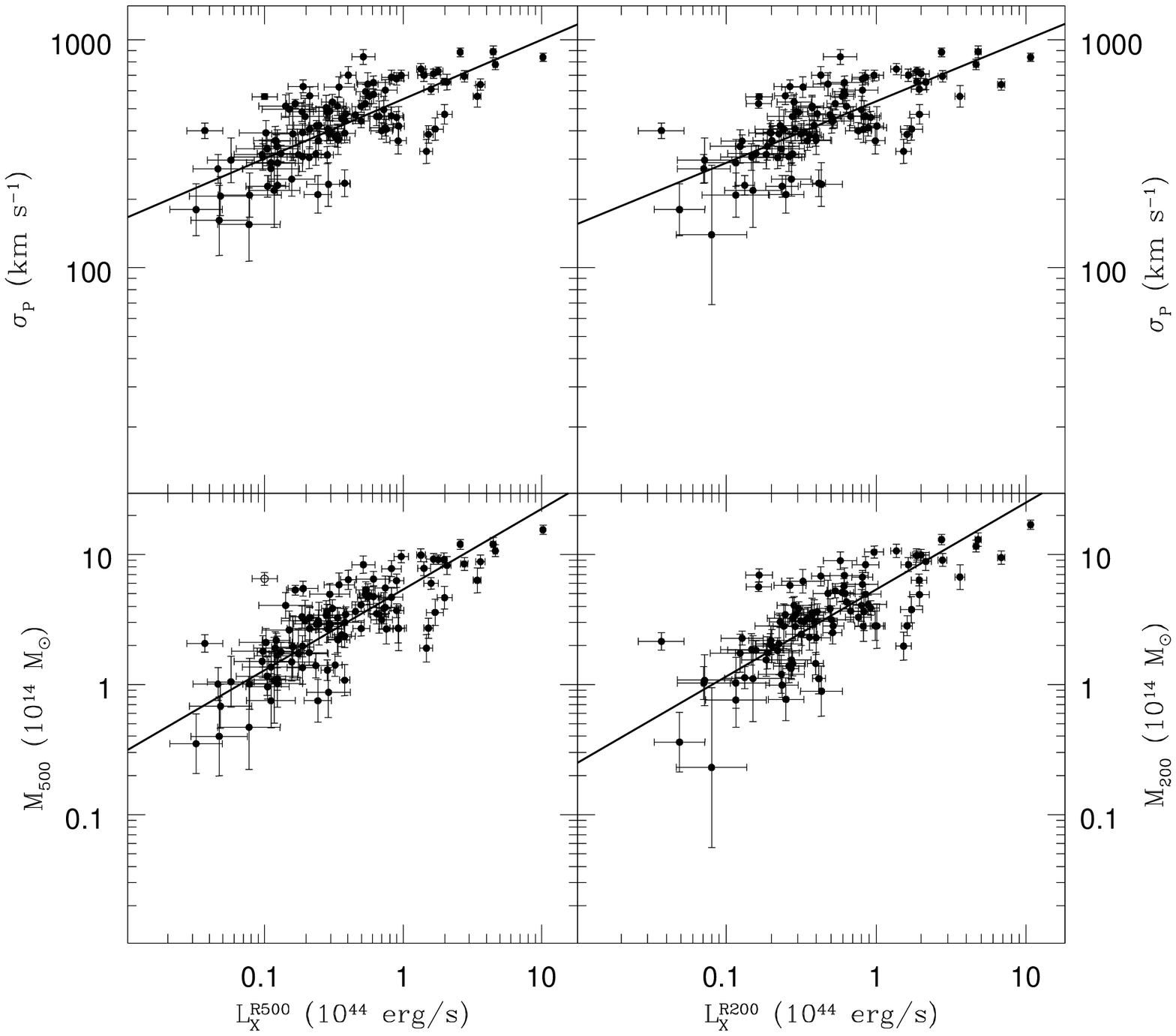}
\end{center}
\caption{The correlation between $\sigma_P$, mass, and X-ray luminosity
(estimated with the ``annulus'' background). The 76 clusters 
with upper limits (UL) for L$_X$ 
are not included in the plot or used for the fitting.}
\label{fig:richscalerelations}
\end{figure*}

Next we compare the X-ray luminosity estimated in \citet {lop09} to
$\sigma_P$ and mass. As described in paper I, we computed the X-ray
luminosity\footnote{Due to a bug in the code used to estimate 
the X-ray luminosity and its 
associated error, the background was not properly determined in paper I. That
resulted in slightly different values for L$_X$. However, the errors of L$_X$
were severely underestimated. We fixed the code and updated the values
of L$_X$ and its error (for all the three backgrounds) in the current paper.
The information given in Table 5 of paper I is now replaced by the data 
available in Table A1, in the appendix of the present paper.}, from RASS, 
using three different background estimates, termed
``annulus'' (from a ring surround the cluster), ``boxes'' (100
randomly selected background boxes) and ``frame'' (from the whole
frame in which the cluster is located). Here we show the
results obtained with the ``annulus'' and ``frame'' backgrounds. 

Figure 5 shows the relation between L$_X$ (``annulus'' background) and
$\sigma_P$ (top panels) and mass (lower panels) , estimated within
R$_{500}$ (left) and R$_{200}$ (right).  We lose seventy-four of the 180
clusters from the extended sample, as their X-ray luminosities are
upper limits. These are mostly poor systems. When establishing the
scaling relations involving L$_X$, upper limits were not used in the
fits. From this plot (and Table 4) we see that the smallest scatter 
is generally found when using R$_{500}$.

We also studied the relations considering the CIRS subsample, with the
comparison of M$_{200}$ and L$_X^{R200}$ shown in Figure 6. It is
particularly important to see whether the slope and mainly the scatter
are reduced for this sample based on richer clusters. As 
found for N$_{gals}$ and L$_{opt}$, the slope also becomes
shallower for the richer systems when considering L$_X$. Note that
the slopes are consistent with the full sample (Figure 5 and Table 4) 
within 1-$\sigma$. The slope of the M$_{200}-$L$_X^{R200}$ relation is 
(0.67 $\pm$ 0.06) for all clusters and (0.52 $\pm$ 0.08) for CIRS 
(see Table 4). The scatters in the 
relations based on the richer clusters are slightly smaller than those of the
whole sample, but still within 1-$\sigma$. The relations based on the 
``frame'' background are similar to the ones obtained with the ``annulus'', 
being the agreement within 1-$\sigma$.  Figures 7 and 8 show the results 
with the ``frame'' background, for the full and CIRS samples,
respectively.

\begin{figure}
\begin{center}
\leavevmode
\includegraphics[width=3.5in]{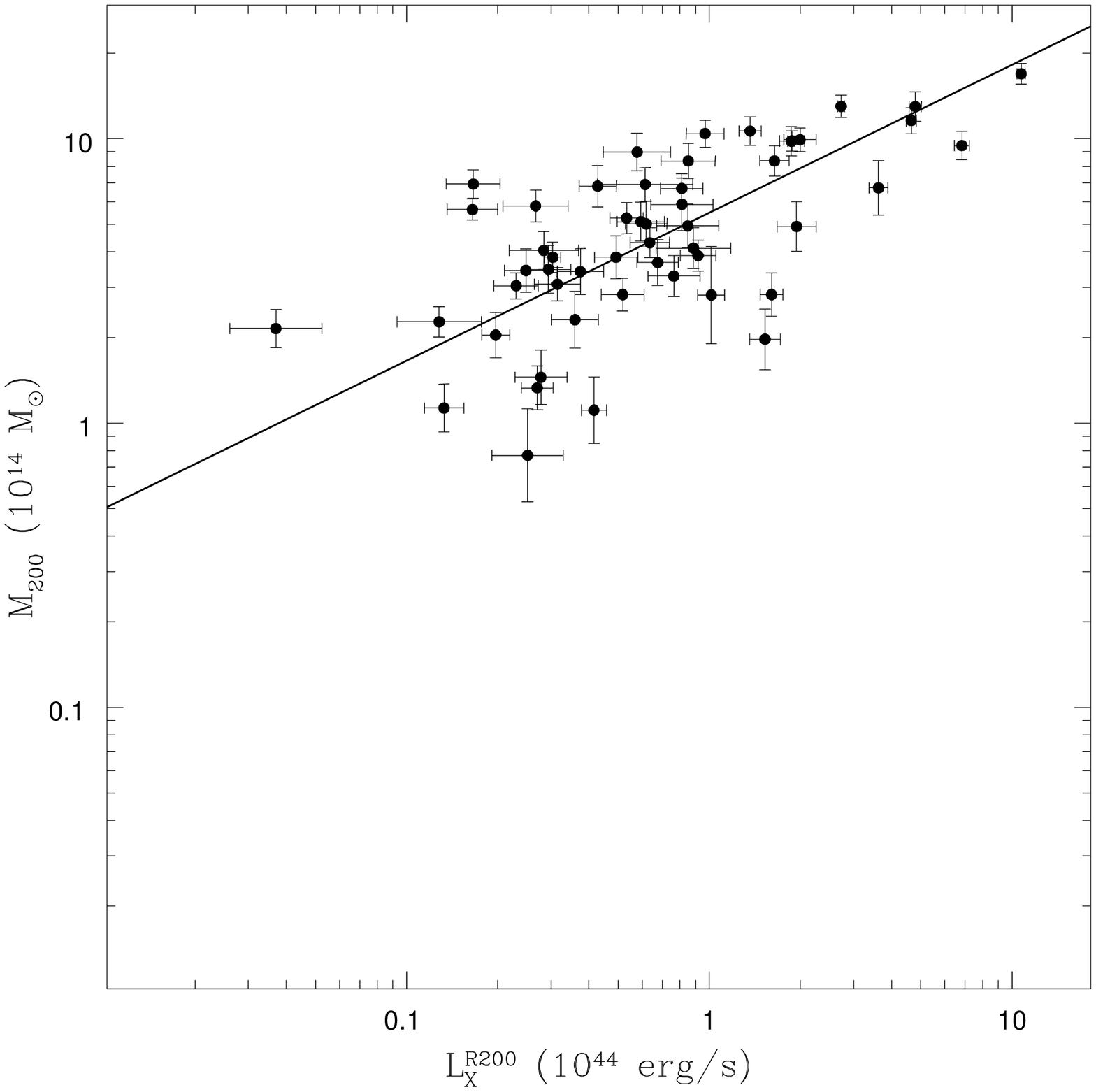}
\end{center}
\caption{Analogous to the previous figure, but showing only the relation 
M$_{200}$-L$_X^{R200}$ for the 53 CIRS clusters.}
\label{fig:richscalerelations}
\end{figure}

\begin{figure}
\begin{center}
\leavevmode
\includegraphics[width=3.5in]{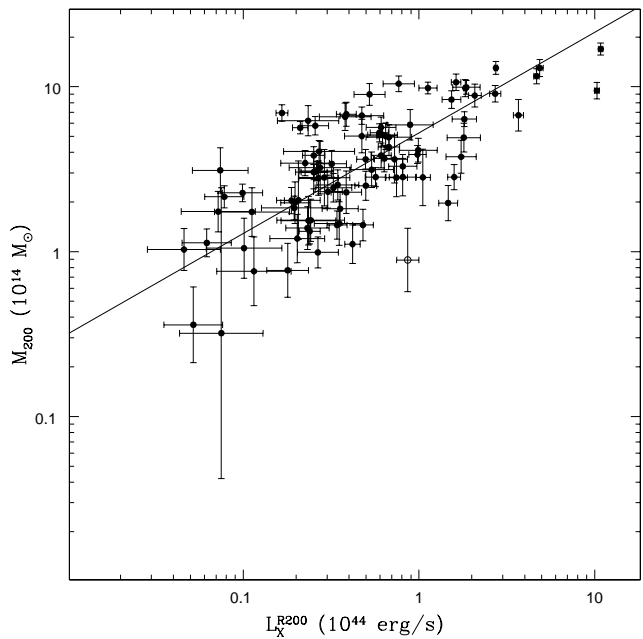}
\end{center}
\caption{Analogous to Figure 5, but showing only the relation 
M$_{200}$-L$_X^{R200}$ for the full sampĺe and with the ``frame'' background.}
\label{fig:richscalerelations}
\end{figure}

\begin{figure}
\begin{center}
\leavevmode
\includegraphics[width=3.5in]{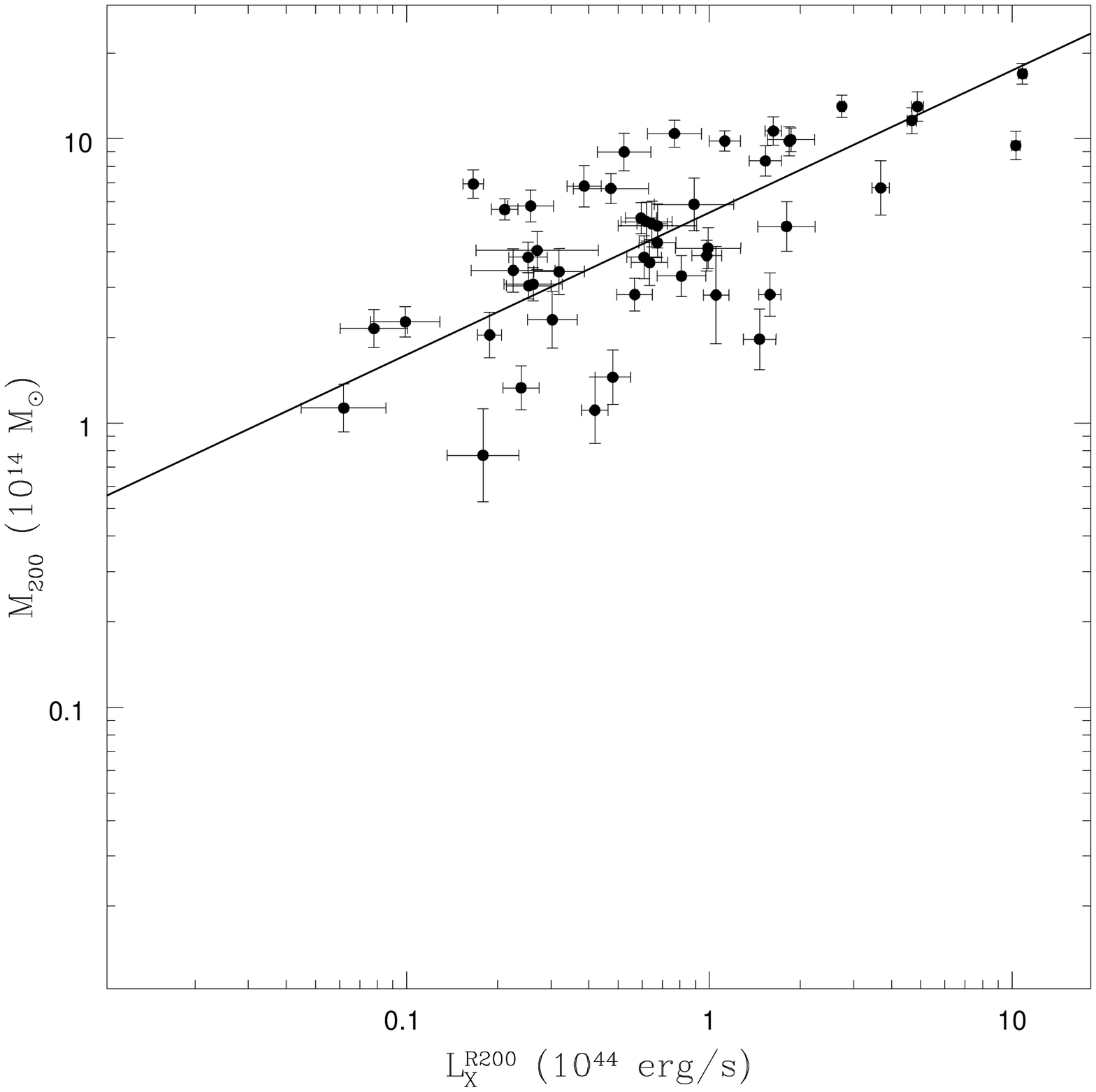}
\end{center}
\caption{Analogous to the previous figure (``frame'' background), but showing 
only the CIRS clusters.}
\label{fig:richscalerelations}
\end{figure}

Tables 4 and 5 summarize the relations involving L$_X$. Table 4
gives the results for the ``annulus'' background, while Table 5 is for 
the ``frame'' background. As in the previous
tables, the first column lists the two parameters
being compared. The background type is
listed in the second column. The sample used is in the third column, while
the remaining columns are analogous to Table 1. In both tables
we list the results for the full sample and for the CIRS systems. 

When comparing the performance of optical parameters (richness or
L$_{opt}$) to L$_X$ as a trace of the cluster mass, we find that the former 
perform a little better than the latter. However, in the worst cases, 
the scatter is consistent within 1.5-$\sigma$ (see Tables 1, 3, 4 and 5). 
The main reason for these minor differences may lie 
in the fact that RASS data is 
shallower than the optical data from SDSS, so that L$_X$ is determined 
less accurately than N$_{gals}$ and L$_{opt}$. It is also important to stress
that the optical properties may be used as reliable mass proxies, in the
same way as L$_X$. This finding corroborates the results from POP05, who found
that L$_{opt}$ gives slightly more accurate results than L$_X$ (also using
SDSS and RASS data).

\subsection{Impact of substructure in the scaling relations}

In \citet {lop09} we described the use of photometric and
spectroscopic data for NoSOCS and CIRS clusters to estimate the
fraction of systems with substructure. Two specific tests have proven
to be very sensitive to the presence of such disturbances in the
galaxy distribution inside a cluster \citep {pin96}. The first test is
the DS, or $\Delta$ test \citep{dre88}, which is a three dimensional
test. The second is a two dimensional test, called the symmetry or
$\beta$ test, introduced by \citet {wes88}.  Detailed descriptions of
both tests can be found in \citet {pin96} and \citet{lop09}. For both
tests, the significance level is determined with the aid of Monte
Carlo simulations. We set our significance threshold at 5$\%$.

The substructure tests are only applied to clusters with at least five
galaxies within the aperture being considered. In paper I we showed
that when using an aperture of R$_{200}$ the rate of clusters showing
significant signs of substructure is $\sim 21 \%$ for both the
$\Delta$ and $\beta$ tests. The latter can be applied to galaxies with
spectra or to the photometric data alone, while the former test
requires redshifts. If we apply the $\beta$ test to the galaxies
within a fixed metric aperture (1.5 h$^{-1}$ Mpc) the rate of clusters
with substructure rises to $\sim 35 \%$, in line with \citet{lop06}, who
estimated substructure from two dimensional optical data for 
$>$ 10,000 clusters.

\begin{figure}
\includegraphics[height=0.50\textwidth,angle=0]{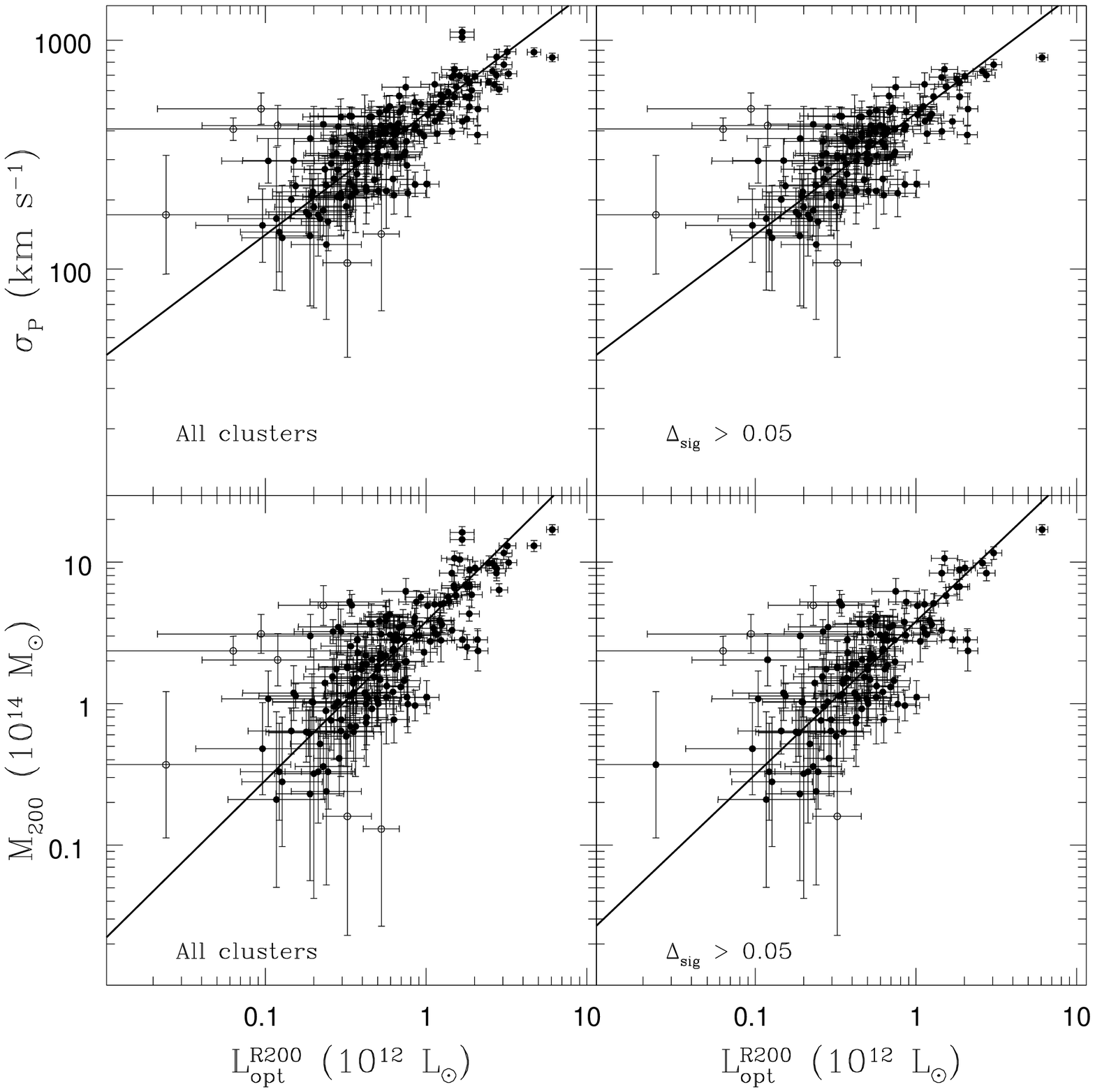}
\caption{The relations $\sigma_P$-L$^{200}_{opt}$ (top panels) and
M$_{200}$-L$^{200}_{opt}$ (bottom panels) are shown for all clusters
(left) and systems without significant evidence for substructure
(right).  We apply the Dressler-Schectman (or $\Delta$) substructure
test to the spectroscopic sample of 170 clusters with at least 5
galaxies (with spectra) within R$_{200}$. Of these, 39 ($\sim$ 23\%)
clusters shown strong signs of substructure.}
\label{fig:zspeczphot}
\end{figure}

\begin{figure}
\includegraphics[height=0.50\textwidth,angle=0]{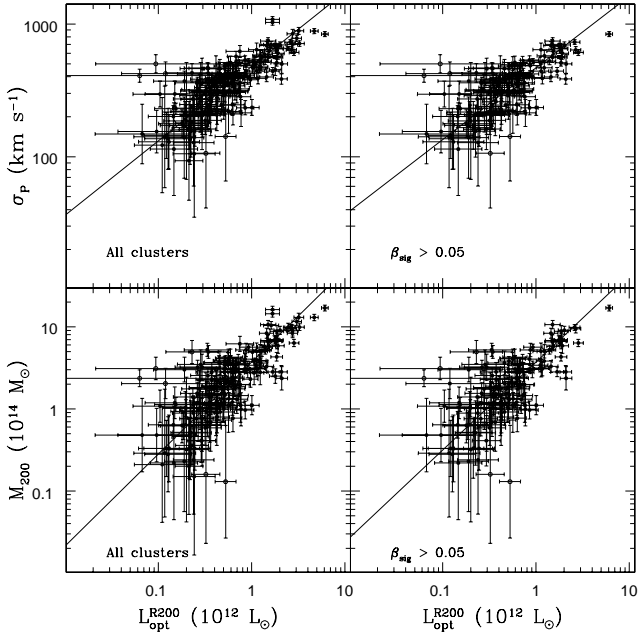}
\caption{Analogous to the previous figure, except for the data set  and 
substructure test employed. Now we use the symmetry (or $\beta$) substructure 
test applied to the photometric data (m$^*-1 \le m_r \le $m$^*+1$ ) for the 179 
clusters with at least 5 galaxies within R$_{200}$. This is a 
two-dimensional (2D) test. Out of the 179 clusters, 43 ($\sim$ 24\%) 
systems are found to have substructure.}
\label{fig:zspeczphot}
\end{figure}

\begin{figure}
\includegraphics[height=0.50\textwidth,angle=0]{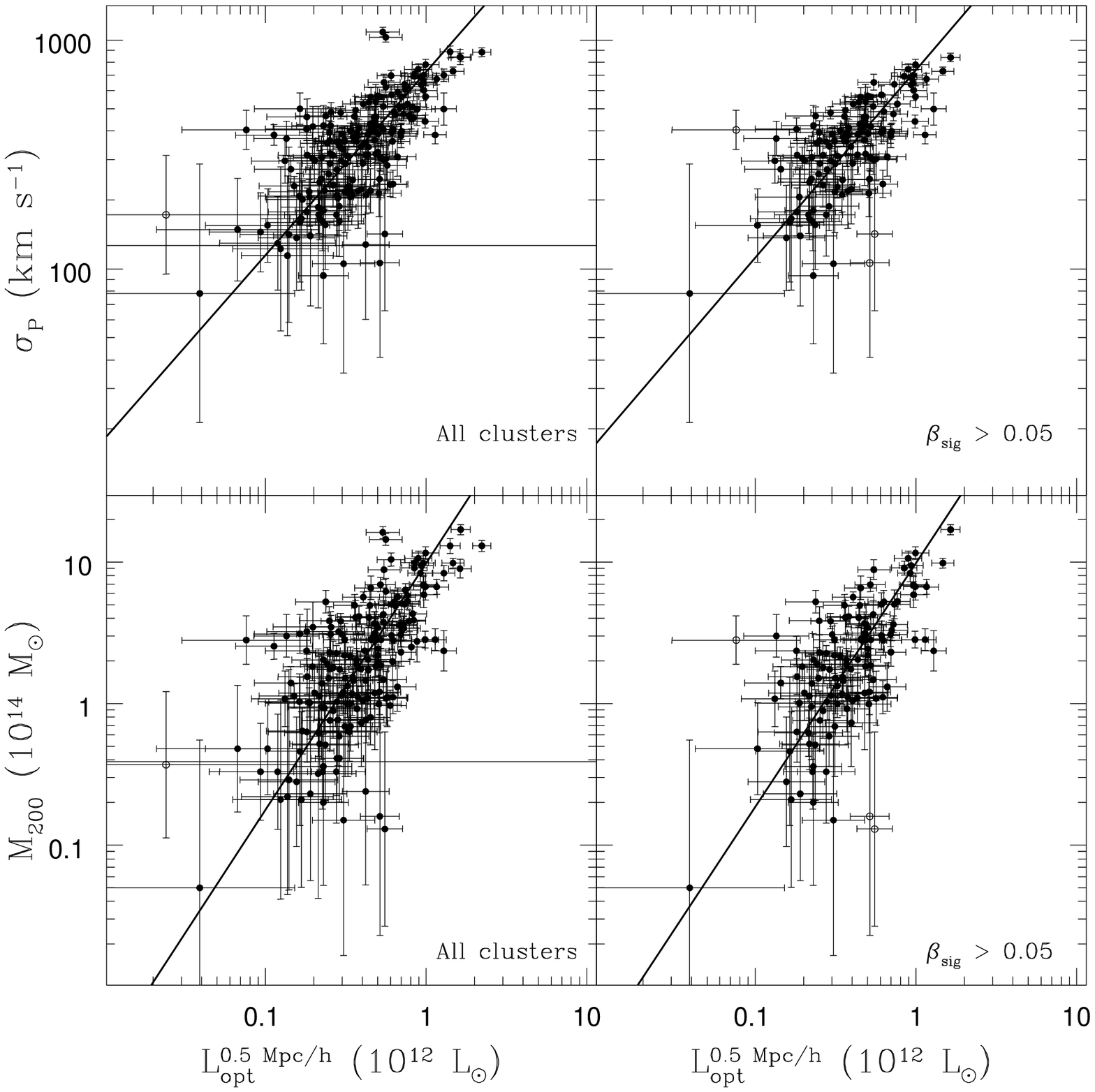}
\caption{Analogous to the previous figure, except for the sample of galaxies 
considered. The $\beta$ test is used again. However, we consider the values 
of L$_{opt}$ computed within 0.5 h$^{-1}$ Mpc and the substructure results 
obtained within 1.5 h$^{-1}$ Mpc. The choice of optimal sizes of the fixed 
aperture for counting galaxies and the way we estimate substructure are 
both explained in \citet {lop06}. The $\beta$ test is applied to all 183 
clusters of the NoSOCS and CIRS sample. We find that 64 ($\sim$ 35\%) 
clusters show signs of substructure.}
\label{fig:zspeczphot}
\end{figure}

In this work we do not intend to investigate substructure for each
cluster individually. Our goal is to check the possible effect of
substructure on the scaling relations connecting M$_{200}$ to
N$_{gals}$ and L$_{opt}$.  Four different cases are considered.  First
we apply the $\Delta$ test to the 170 clusters with at least five
galaxies with z$_{\rm spec}$ within R$_{200}$. Second, we apply the
$\beta$ test to the same data set. Third, the $\beta$ test is applied
to the photometric galaxy distribution within R$_{200}$. We consider
all galaxies with m$^*-1 \le m_r \le $m$^*+1$ (see \citealt {lop06}
for details) with no restriction regarding z$_{\rm spec}$. Fourth, we
apply the $\beta$ test to all galaxies within 1.5 h$^{-1}$ Mpc of each
cluster (the same photometric range is enforced).

The results for the cases listed above are summarized in Table 6. In
this table we always show first the results for the full sample and
then the ones for clusters without substructure. The first six lines
give the results considering the first two cases mentioned above.
Lines 7 to 10 list the results for the third case, and the last four
lines summarize the results for the fourth case. The first column
lists the two parameters of the scaling relation (and the aperture
adopted for computing N$_{gals}$ and L$_{opt}$). The second column
shows the substructure test (if any) used to remove clusters with
substructure. Note that the aperture used for estimating substructure
is R$_{200}$ in the first three cases and is 1.5 h$^{-1}$ Mpc for the
fourth. The sample used is listed in column 3. The meaning of the
remaining columns is the same as in Table 3.

Figures 9, 10, and 11 show the results for the first, third and fourth 
cases discussed above. In each figure the relation between $\sigma_P$ (top) 
and M$_{200}$ (bottom) to L$_{opt}$ is displayed for all clusters (left) and 
for the substructure-free systems (right). From Table 6 and Figures 9 and 10 
we see that the fitting parameters and scatter of the scaling relations, 
depend very little on the presence of substructure. The intercepts and slopes
are always consistent within 1-$\sigma$. It is important to note 
that the 2D and 3D tests indicate similar fractions of clusters with 
substructure, and the scaling relations obtained for the substructure-free 
systems are nearly the same as those for the full sample, 
regardless of whether the test used to detect substructure is 2D or 3D. 

From Figure 11 we see that even considering the full photometric data 
within a fixed radius, instead of only cluster members within R$_{200}$, the 
results are qualitatively the same. Due to the different apertures the 
scaling relations are different from the two previous figures.
However, we still find the relations
to be insensitive to the exclusion of clusters with substructure. The 
scatter of the relations (with or without substructure) shows a larger 
difference than for the other cases above, but it is still within 1-$\sigma$.
These results contradict Figure 16 of 
\citet {lop06}, where a strong segregation in the T$_X$-N$_{gals}$ relation 
was caused by the presence of substructure in some clusters. Several reasons 
may be responsible: 1) T$_X$ was taken from the literature, so the 
estimates may be very heterogeneous; 2) T$_X$ was not available for the poorer 
clusters (N$_{gals} <$ 20); and 3) the small sample size used in that work 
makes the results sensitive to the exclusion of a few points. 

We repeated the analysis presented in Figure 11 but only for clusters 
with N$_{gals} \ge$ 20 and found that the scatter of the relations 
decreased only by 5\%. However, the slope exhibited a significant 
variation ($\approx 27 \%$). So, the strong segregation present in the 
T$_X$-N$_{gals}$ relation of \citet {lop06} may be explained by the small 
sample size and the exclusion of poor systems (N$_{gals} <$ 20). In 
\citet {lop09}, we checked that the T$_X$ values derived from BAX represent 
a consistent data set, which implies that the main source of slope variation 
detected is the absence of low mass systems.

\subsection{Comparison of scaling relations derived with optical and X-ray mass estimates}

As described in \citet {lop09} we have searched BAX ({\it Base de Donn\'ees 
Amas de Galaxies X}, http://bax.ast.obs-mip.fr/) for counterparts to the 183 
clusters used in this work. The search was restricted to objects at $z < 0.12$ 
with X-ray temperature measures. We found 282 clusters in BAX, of
which 21 are common to our sample. For these clusters, we employed the 
M$_{200}$-T$_X$ relation given by equation 3 of POP05 to estimate 
masses. Our goal is to check if the mass calibration 
gives the same results when using either the optical or X-ray masses. 
Of the 183 NoSOCS plus CIRS systems, temperatures are generally
available only for the more massive clusters, with $\sigma_P >$ 400 km/s or 
M$_{200} >$ 10$^{14}$ M$\odot$. 

In \citet{lop09} we show that the temperature values selected from BAX 
represent a consistent data set. There, we compare the values listed in BAX 
to the ones available in RD06, finding that most systems agree 
within 10\%. We also check that the masses obtained from the dynamical 
analysis of the optical data generally agree within 40\% of the estimates 
derived from the M$_{200}$-T$_X$ relation. The exceptions, most of times, are 
clusters affected by substructure. When using the optical masses derived 
with the caustic technique (RD06) the agreement is not as good, 
with the optical masses being lower than the X-ray values in the low-mass 
regime. 

In this section, we check if the scaling relations 
obtained in this work are similar to the ones using mass derived from the 
tight connection to T$_X$. As the velocity dispersions and masses obtained 
with the caustic technique are provided by RD06 we also compare 
our results to those based on their estimates. Figures 12 and 13 
show the M$_{200}$-L$_{opt}$ and 
M$_{200}$-L$_X$ relations, respectively. L$_X$ is the value derived in this 
work with the ``frame'' background, using RASS data. Both 
figures consider M$_{200}$ as determined from 
T$_X$ (equation 3 of POP05). These results are summarized in 
the first four lines of Table 7, where we also list the fit parameters for 
the M$_{200}$-N$_{gals}$ relation and for M$_{200}$-L$_X$ (considering 
the ``annulus'' background).

\begin{figure}
\begin{center}
\leavevmode
\includegraphics[width=3.5in]{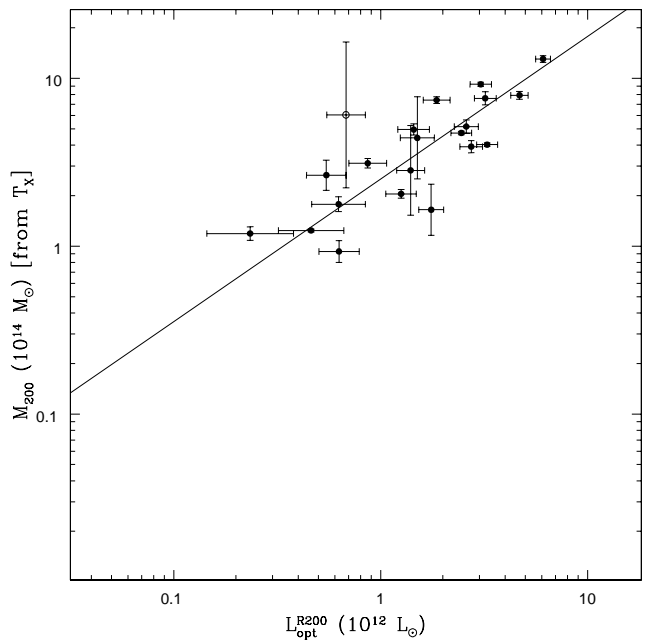}
\end{center}
\caption{The relation between mass estimated from X-ray data and optical 
luminosity.}
\label{fig:richscalerelations}
\end{figure}

\begin{figure}
\begin{center}
\leavevmode
\includegraphics[width=3.5in]{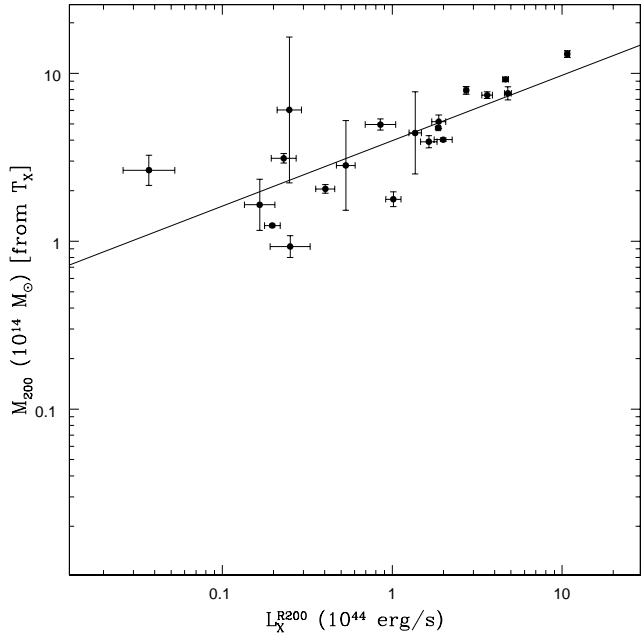}
\end{center}
\caption{The relation between mass estimated from T$_X$ and X-ray 
luminosity. L$_X$ is the one determined in this work, considering the 
``frame'' background.}
\label{fig:richscalerelations}
\end{figure}

From inspection of this table, and considering the range of clusters
sampled, the results are in very good agreement to those based on the
virial mass and restricted to the richer clusters. In Tables 1, 3, 4 and 5 
the CIRS sample represents the richer clusters. Although the sample with 
masses derived from T$_X$ (Table 7) is less than half of our rich 
sample, we conclude that using the optical or X-ray mass leads essentially 
to the same scaling relations. In particular, the slopes are always concordant
within 1-$\sigma$. It is imperative to compare the results
in Table 7 to those of the rich samples of the previous tables, which
have approximately the same mass range.

Note also that the scatters of the relations
in Table 7 are generally smaller than what we found with the virial masses.
However, the results agree within 1-$\sigma$. The improvement is more 
pronounced only for the relations involving
L$_X$, reaching half the scatter for the ``frame'' background (but still
within 2-$\sigma$). These more accurate results are due to the smaller
sample size of the relations based on the mass obtained from T$_X$ (only 21
clusters). The most important conclusion from this exercise is that 
N$_{gals}$, L$_{opt}$ and L$_X$ can be used for mass calibration and the 
results are independent to the way mass is estimated, from optical (virial
analysis) or X-rays (using the M-T$_X$ relation). 

We decided to investigate if these conclusions hold for the parameters derived 
from the caustic technique (RD06). Figures 14, 15, 16, and 17 show the 
M$_{200}$-N$_{gals}$, M$_{200}$-L$_{opt}$, M$_{200}$-L$_X$ and 
$\sigma_P$-L$_{opt}$ relations, considering mass and $\sigma_P$ obtained 
with the caustic technique. These parameters are provided for all CIRS 
clusters in RD06. Richness, optical and X-ray luminosities are 
estimated in the present work. These 
relations, as well as $\sigma_P$-N$_{gals}$ and $\sigma_P$-L$_X$, are 
summarized in the last eight rows of Table 7. 

When comparing these relations to the results in Tables 1, 3, 4 and 5
(only for the CIRS systems) the normalizations show better agreement
than above.  This is due to the fact that all the comparisons use only
the CIRS sample. The slopes agree within 2-$\sigma$ for the
relations involving $\sigma_P$ and within 1-$\sigma$ for those
regarding M$_{200}$. However, this only happens because the uncertainties
in the fit parameters of the relations involving mass are much larger
than the ones regarding $\sigma_P$. The uncertainty in the slopes of
the relations using the caustic mass is also larger than what we found
with the virial mass, by as much as a factor of three. So, although
the relations based on our masses (or the X-ray derived masses) agree
within the errors to those obtained with the caustic method, the
nominal values are very different. When using the caustic parameters,
the relations about $\sigma_P$ have smaller slopes compared to
ours. The opposite is true for the relations regarding M$_{200}$.

We also find that the relations based on the caustic values are noisier 
(with larger scatter) than ours, especially for results 
involving mass. This is easily seen in Figures 14-17 and Table 7. For 
instance, for the M$_{200}$-L$_{opt}^{R200}$ relation the scatter in mass at
fixed luminosity is (0.74 $\pm$ 0.18) when using the CIRS values, but only 
(0.36 $\pm$ 0.06) for the virial masses determined in the current work. 

Such findings are explained by the fact that our velocity dispersions
are in good agreement with those of RD06, as shown in paper
I. However, the same is not true when comparing masses. Those masses
estimated from T$_X$ also disagree with the caustic values. RD06
claim concordance between the caustic results and their virial masses
(as well as the X-ray estimates). However, comparing our masses and
the caustic values (RD06) to the X-ray estimates, we find a better
correlation with our values. Therefore, we conclude that our estimates
better represent the cluster potentials, as discussed in Section 6 of
paper I.  There, we also argue that the interloper removal procedure
should not be responsible for the different results (as the velocity
dispersions are similar between our work and RD06).

The culprit for the different mass estimates lies in the discrepant values of 
R$_{200}$, which in RD06 are derived from the caustic mass profile.
In paper I we estimate R$_{200}$ as a by-product of the virial analysis and
the assumption of an NFW profile \citep{nar97}. This profile is well
matched to the caustic mass profile for only half of the CIRS sample
(RD06), which may explain the different results and the larger
scatter for the caustic based scaling relations. Note also that RD06
use R$_{200}$ determined by the caustic to compute the virial masses. 
That choice
helps improving the agreement they find between the caustic and virial masses.
The conclusions we reached in paper I \citep{lop09} are corroborated by the 
scaling relations shown above.

\begin{figure}
\begin{center}
\leavevmode
\includegraphics[width=3.5in]{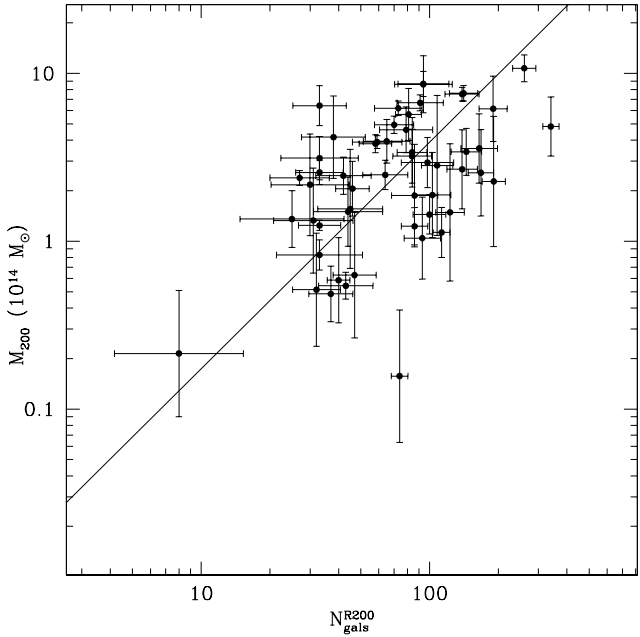}
\end{center}
\caption{The connection between the caustic mass (RD06, estimated 
from the optical data) and richness. The relation is exhibited for the 53 
CIRS clusters.}
\label{fig:richscalerelations}
\end{figure}

\begin{figure}
\begin{center}
\leavevmode
\includegraphics[width=3.5in]{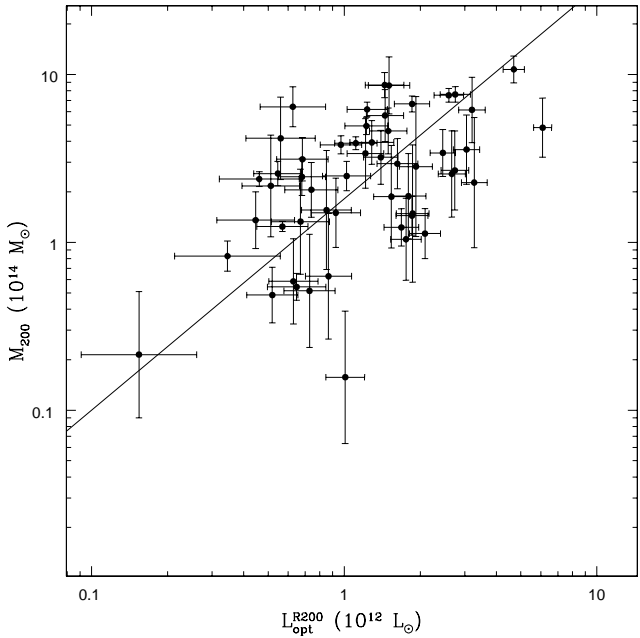}
\end{center}
\caption{Analogous to previous figure, but showing L$_{opt}$.}
\label{fig:richscalerelations}
\end{figure}

\begin{figure}
\begin{center}
\leavevmode
\includegraphics[width=3.5in]{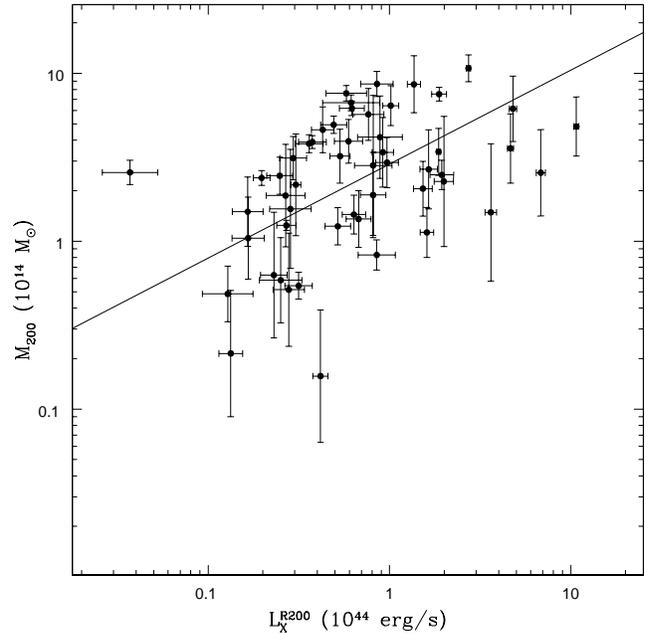}
\end{center}
\caption{The relation between the caustic mass (RD06) and X-ray 
luminosity. L$_X$ is obtained with the 
``annulus'' background.}
\label{fig:richscalerelations}
\end{figure}

\begin{figure}
\begin{center}
\leavevmode
\includegraphics[width=3.5in]{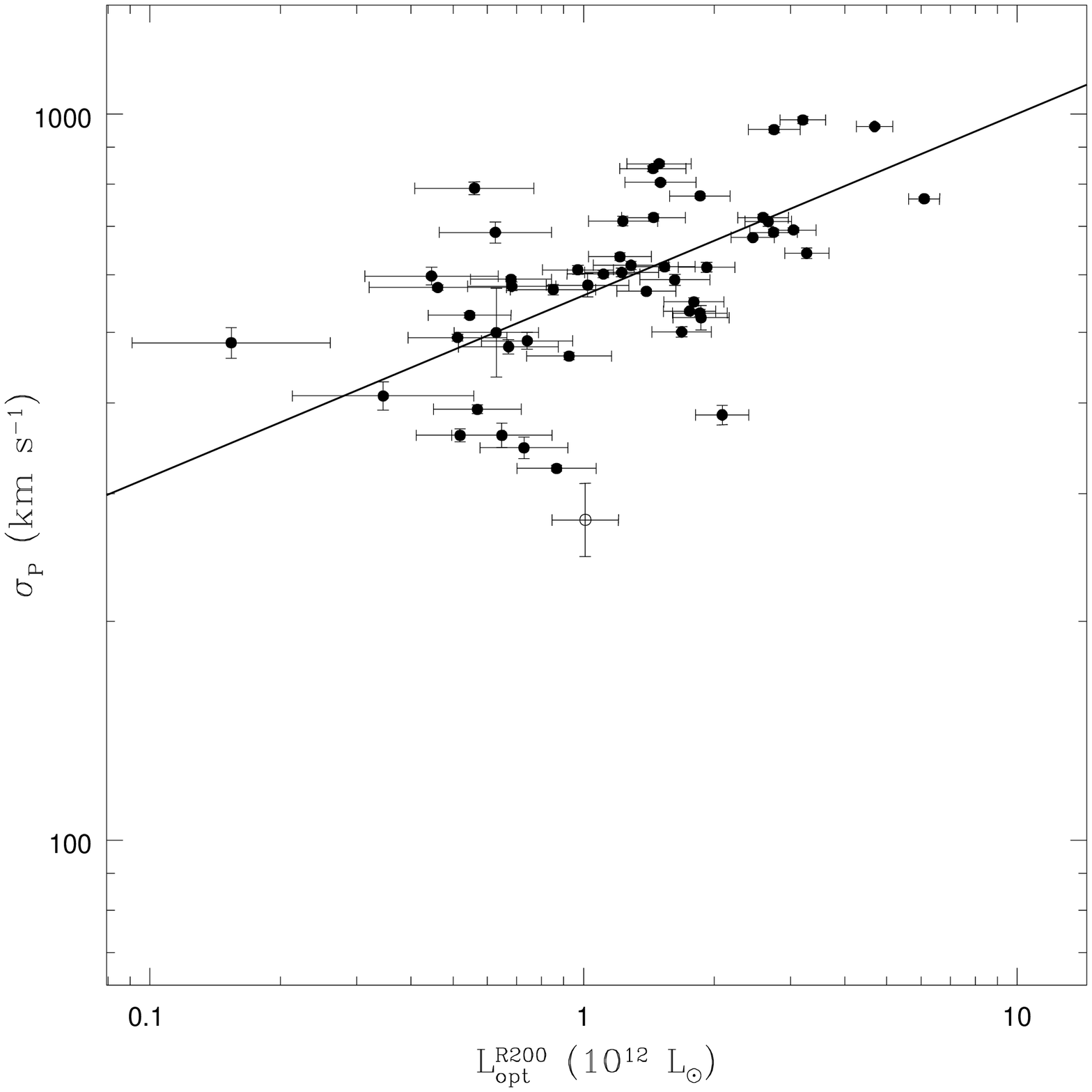}
\end{center}
\caption{The relation between velocity dispersion estimated with the caustic 
technique and optical luminosity. The relation is shown for the 53 CIRS 
clusters.}
\label{fig:richscalerelations}
\end{figure}

\subsection{The mass-to-light ratio}

In this section, we present the mass-to-light ratio (M/L) of clusters of 
galaxies. That has been used before for estimating the mass density of 
the universe. The typical value of M/L for rich clusters is 
M/L $\simeq$ 300 h M$_{\odot}$/L$_{\odot}$ \citep{bah02}.
Previous works have shown that M/L increases with the size 
of the bound systems, from
galaxies to clusters. The dependence of
the cluster M/L ratio on cluster mass can be parametrized by a relation like
M/L $\propto$ M$^{\alpha}$, with ${\alpha} \simeq 0.20$ \citep{bah02, pop05}. 

In Figure 18, we show the mass-to-light ratio obtained in this work.
We recomputed L$_{opt}$ considering the interval of
m$^*$-5 to m$^*$+5. Using the values obtained in 
paper I and considering the same range used for the richness estimation 
(m$^*$-1 to m$^*$+2), would lead to M/L values that 
are too high. We do not consider 
an incompleteness correction to the luminosity function (LF) for 
galaxies fainter than m$^*$+5 as the correction is around
5\% and uncertainties in the slope at the faint end of the LF could bias the 
L$_{opt}$ values. So, integrating the LF to infinity does not significantly
change L$_{opt}$. The range we considered is already deep enough (being
five magnitudes fainter than m$^*$). In Figure 18, the solid line shows 
the relation listed in Table 3 for ALL clusters, while the other two are based 
in \citet{pop05}. The dotted line considers 
their ``optical'' sample (69 clusters with optical masses), while the dashed 
line is for their ``enlarged'' sample, with 102 clusters comprising the
optical sample plus clusters with masses estimated from the M-T$_X$ relation. 
All the lines are normalized to an M/L ratio of 200.

As we can see from Figure 18 our results show good consistency to previous
findings, reinforcing the dependence of the M/L ratio to the cluster scale.
Note that the power of the relation we find is $\alpha = 0.10$, smaller than 
the value of 0.19 from \citet{pop05} (``optical'' sample; dotted line). 
However, both results are consistent within 1-$\sigma$ and we find nearly the 
same slope as represented by their ``enlarged'' sample ($\alpha = 0.09$; 
dashed line). The mean M/L ratio we find is $\sim 210$, in excellent 
agreement to previous studies \citep{bah02}. In a future work we plan 
to investigate the LF function of clusters in detail and in conjunction to 
the mass-to-light ratio and its dependence on cluster mass.

\begin{figure}
\begin{center}
\leavevmode
\includegraphics[width=3.5in]{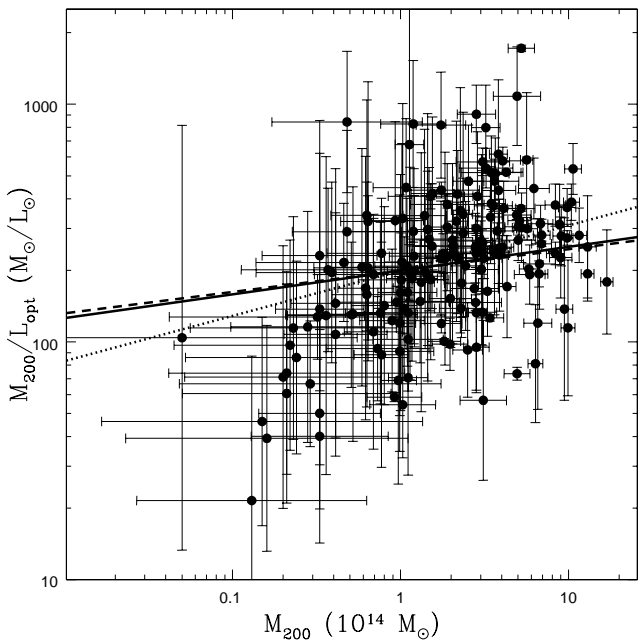}
\end{center}
\caption{The mass-to-light ratio for the 127 NoSOCS plus 53 CIRS clusters.
The solid line shows the relation listed in Table 3 for ALL clusters. The 
other two results shown are from \citet{pop05}. The dotted line considers 
their ``optical'' sample, while the dashed line is for their ``enlarged'' 
sample. All the lines are normalized to a M/L ratio of 200.}
\label{fig:richscalerelations}
\end{figure}

\section{Comparison to the literature and discussion}

The results in Tables 1, 3, 4 and 5 indicate a good agreement with
most findings in the literature. In Table 8 we summarize our main
results, considering all clusters, or only the CIRS systems, as well
as the findings of other authors. In particular, there is good consistency 
between the slopes and scatter of POP05 (also based on SDSS 
and RASS) with our results, in the
worst cases within 1.5-$\sigma$. RD06 claim to find good consistency
with POP05. However, as seen in the previous section that is at the
cost of larger errors for the fit parameters and much noisier
relations. Note that our results based solely on CIRS (the most
massive systems) always have smaller slopes (an issue we discuss
further below).  Using an enlarged data set \citet {pop07} found relations 
consistent to POP05.

The works of POP05 and RD06 are the easiest comparison to ours, since
they are also based in SDSS data. However, other results, based on
different surveys and wavelengths can be informative. \citet {lin04}
found M$_{200} \propto$ L$_{K,R200}^{1.22}$ and M$_{200} \propto$
N$_{gal,R200}^{1.15}$, using Two Micron All Sky Survey (2MASS)
data. Our slopes agree within 1.5 and 1-$\sigma$,
respectively. Note that we consider their results after excluding the
BCG, as they claim better agreement on the powers of the two
relations. \citet {koc03} found M$_{200} \propto$ N$_{*,666}^{0.91}$,
also using the 2MASS K-band, a result that agrees with ours within
1.5-$\sigma$. These findings corroborate the conclusion of \citet
{pop07}, who reports no significant difference among results in
different SDSS bands. So, even considering the 2MASS K-band it seems
that the results are similar.

On what regards other works using L$_X$ as a mass tracer, we find that the 
slopes of \citet{pop05, sta06, vik08} and \citet{ett04} are 
consistent to ours. The scatter is consistent (within 1-$\sigma$) to 
the works of \citet{pop05} and \citet{sta06}, while the scatter from 
\citet{vik08} and \citet{ett04} are smaller than ours (consistent only 
within 3-$\sigma$). Note that the first two works above \citep{pop05, sta06} 
are based in the RASS data, while the last two use the Chandra satellite.
We find a very good agreement (within 1-$\sigma$) between the normalization we 
measure and the results of \citet{pop05}, for the M-L$_X$ relation. Differences
in the normalization to the other works above might be due to the differences
in the cluster samples used, which may span different mass/luminosity regimes.
The comparison of the different selection functions is very hard and is not
our goal in this work.

The small differences with POP05 still deserve a few comments. The
scatter in M$_{500}$ at fixed L$_{opt}^{R500}$ is similar to ours (consistency
within 1.5-$\sigma$). However, a closer look at the
upper left panel of our Figure 4 and their Figure 5 shows that we
have many more clusters at M$_{500} < $10$^{14}$M$_{\odot}$ than they
do.  So, their results are based on a data
set showing a cloud of points at M$_{500} > $10$^{14}$M$_{\odot}$ and
a few scattered systems below this value. Nonetheless, that small data
set in the low mass regime is sufficient to put their results and ours
at the same level. One odd aspect of their plot is the diffuse
distribution of the high mass systems, showing high scatter in this
mass regime. Our relations always show a reduced scatter towards
higher mass. Their poor sampling at low masses can also be seen in
Figures 11 and 13 of their work, where we note a deficit of clusters
below $\sigma_P =$ 400 km/s.

Considering this sampling issue  in POP05 it is natural 
to compare their results with the ones we obtained using only CIRS (which also 
has mostly clusters with $\sigma_P >$ 400 km/s). Figure 3 shows the relation 
between $\sigma_P$ and L$_{opt}^{R200}$. It is remarkable how low the scatter 
is in this figure. However, the most interesting 
point is that the relations are slightly flattened for the CIRS clusters, 
when compared only to NoSOCS or all systems. So, it seems that the scaling 
relations have different behaviors in 
different mass regimes. They look steeper 
for the low-mass systems, becoming flatter for the high-mass clusters. 
Using CIRS we note that the agreement to POP05 is as good as before,
despite the fact the slope is now smaller. 

\citet{pop07}  created a larger sample by adding 130 
Abell clusters with confirmed 3D  
overdensities in SDSS. Although this sample is larger than ours (217 clusters 
compared to 180) it still seems that there are few systems below 
$< $10$^{14}$M$_{\odot}$ (see their Figure 9). Most  confirmed Abell
overdensities are rich systems, which explain the figure (note that the 
open symbols are the X-ray systems, which contain the lower mass objects). 
The main conclusions drawn from these comparisons are: (i) the slopes of the 
scaling relations become flatter for high-mass samples, but the differences
are within 1-$\sigma$; (ii) although the samples of POP05 and
\citet{pop07} show fewer systems with low mass, their results are consistent 
with ours. That is due to the inclusion of a few groups in their sample and
to the small variation in the slope with mass. 
These results indicate that sample
incompleteness has little effect on the scaling relations.

For the relations involving L$_X$ we find the same trend as 
above, with the high mass systems showing flatter relations. However,
some clusters were not considered in the fits due to having only upper limits 
in the X-ray luminosity estimates. This eliminates points that generally 
have L$_X <$ 10$^{43}$ ergs/s in Figure 5 (lower right panel). These points 
would be located above the derived fit and would thus make the relations a 
little flatter. Our results with all clusters could be biased 
due to missing some low mass systems with L$_X \sim$ 10$^{43}$ ergs/s, 
making the relations steeper 
than it would really be. However, considering the conclusions reached above,
we might expect that incompleteness (especially in the low mass regime) has
a minor impact in the relations. We also note that the results within 
R$_{500}$ generally are the ones with smallest scatter when considering L$_X$. 

A result that deserves some comments is the third test shown in Table
2 (rows 5 and 6) where we consider all clusters at $z \le$ 0.25,
considering that at $z >$ 0.10 the spectroscopic survey of SDSS is no
longer complete. The main effect exhibited is the flattening of the
scaling relations. That happens because N$_{gals}$ and L$_{opt}$ are
still well determined (as they rely on the photometric data), but
$\sigma_P$ and mass are not (see discussion in paper I). If the percentage
of clusters without enough spectra is large, the impact
in the scaling relations may be important. So, the safest approach is to
consider only objects that have complete spectroscopic sampling (as
done here and by RD06).

Another important issue concerns the interloper removal procedure (see
\citealt {woj07}).  As discussed in $\S$4 of \citet {lop09} we
performed several tests to optimize the ``shifting-gapper''
technique. The final code seems to work well for systems of all
masses, as can be seen from the phase-space diagrams \citep
{lop09}. POP05 considers the method of \citet{kat04} (see also
\citealt {den96}), which also combines the position and velocity
information, but makes assumptions regarding the dynamical state of
the cluster. RD06 use the caustic technique to determine the mass
profile of the clusters, which involves removal of interlopers in
underdense regions of the phase-space diagram. \citet {woj07} applied
several different techniques for interloper removal to real and
simulated data. They conclude that some methods have poor performance
compared to others, but none of the methods cited above are among the
poor performers. Whatever the approach, it must be sufficiently
restrictive but not overly lax, otherwise many cluster members will be
discarded or few interlopers will be rejected, leading to biased
values of $\sigma_p$ and mass.  While there is no particular reason to
chose one method over another, the procedure should work for a wide
variety of systems (as shown for the method adopted here).  We based
our choice on the simplicity of the method and on the avoidance of
assumptions regarding the dynamics of the cluster \citep {fad96}.

Results from a test with a less rigorous criterion (eliminating fewer
interlopers) are shown in rows 3 and 4 of Table 2.  The comparison of
Tables 1 and 2 reveals no systematic effect on the intercept and
slope of the scaling relations when relaxing the interloper removal
procedure. The scatter of the relations increases slightly, but 
the results are still consistent with our original
relations and to those in the literature. So, we conclude that our
method is robust, rendering similar lists of cluster members than
other methods. This is reflected in the good agreement with the velocity
dispersions of RD06 and the comparison to the scaling relations of
other authors \citep{koc03, lin04, pop05}.

The Halo Occupation Distribution (HOD) relates the number of galaxies within 
the virial radius and the associated mass, expressed as 
N$_{gal,R200} \propto$ M$_{200}^\alpha$. Hierarchical models of structure 
formation predict that the number of subhalos within a system is directly 
proportional to the mass of the parent halo. In other words, $\alpha = $ 1. 
However, different mechanisms (such as a decreasing efficiency of star 
formation, or an increased merger rate) would 
imply a decreasing number of galaxies per given mass in higher mass halos, 
agreeing with $\alpha < $ 1 (see discussion in \citealt {pop07}). Here, we do 
not intend to investigate the HOD. Our main goals are to provide calibrators 
to the cluster mass and compare those to determine which one traces mass more 
accurately. However, it is worth mentioning that the connection between mass 
and richness we found is indeed consistent with $\alpha < $ 1, as most recent 
results in the literature. In  a future work we plan to study in detail the
HOD and the factors that could make $\alpha < $ 1. 

Finally, we would like to comment that when comparing the performance of 
N$_{gals}$, L$_{opt}$ and L$_X$ as mass tracers, we find that the optical 
parameters show relations slightly tighter than L$_X$. The scatter of 
M$_{500}$ at fixed N$_{gal}^{R500}$, L$_{opt}^{R500}$ and L$_{X}^{R500}$ is 
(0.43 $\pm$ 0.03), (0.44 $\pm$ 0.04) and (0.47 $\pm$ 0.05), respectively. The 
scatter of M$_{200}$ at fixed N$_{gal}^{R200}$, L$_{opt}^{R200}$ and 
L$_{X}^{R200}$ is (0.46 $\pm$ 0.04), (0.49 $\pm$ 0.04) and (0.56 $\pm$ 0.06), 
respectively. In all cases, the scatter and the 
slope of the relations are reduced for high mass systems. POP05 argued that 
the poorer performance of L$_X$ is likely due to the variation in the 
compactness of the galaxy clusters. They also investigated if the scatter 
in the M-L$_{X}$ relation could be due to cooling core effects, but they 
find that this accounts for at most 3\% of the scatter. The low signal 
to noise in RASS (especially for the low-mass systems) can also contribute 
to the higher scatter found in the relations involving L$_X$. However, it is 
important to note that here we quote the errors in the scatter, finding that 
they are all consistent (in the worst case within 1.5-$\sigma$) for the three
mass proxies considered (N$_{gals}$, L$_{opt}$ and L$_X$).

\section{Correlation between optical and X-ray properties}

The interplay between global properties obtained in the optical and 
X-ray regimes are crucial for understanding the complex physics present in
galaxy clusters. In recent years a number of works have been dedicated to 
the comparison of X-ray and optical catalogs or the construction of combined
samples \citep{don01, don02, gil04, pop04, pop05, lop06}. The ability to 
predict the X-ray luminosity or temperature 
from optical parameters, and vice-versa, is 
also important for future surveys, conducted only in one regime. As a 
by-product these comparisons may also reveal unusual clusters (such as 
X-ray underluminous ones), which are very interesting for follow-up studies.

\begin{figure}
\begin{center}
\leavevmode
\includegraphics[width=3.5in]{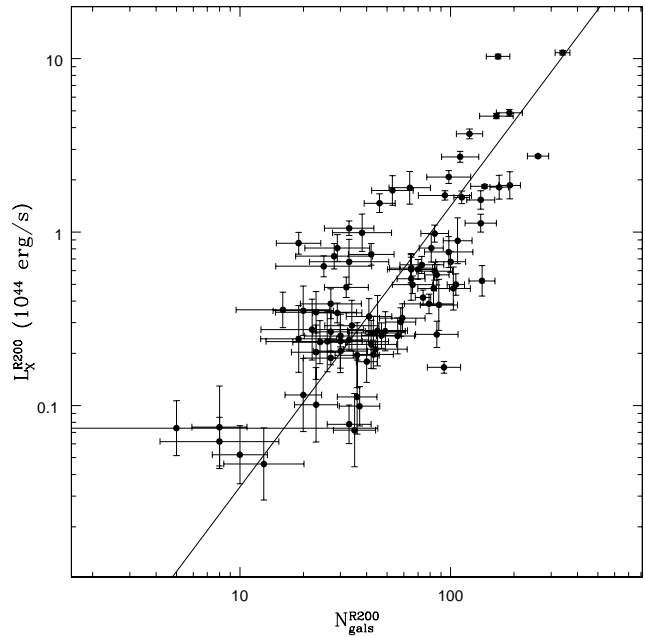}
\end{center}
\caption{Correlation between X-ray luminosity and richness, both 
measured within R200. L$_X$ is estimated with the ``frame'' background.}
\label{fig:richscalerelations}
\end{figure}

\begin{figure}
\begin{center}
\leavevmode
\includegraphics[width=3.5in]{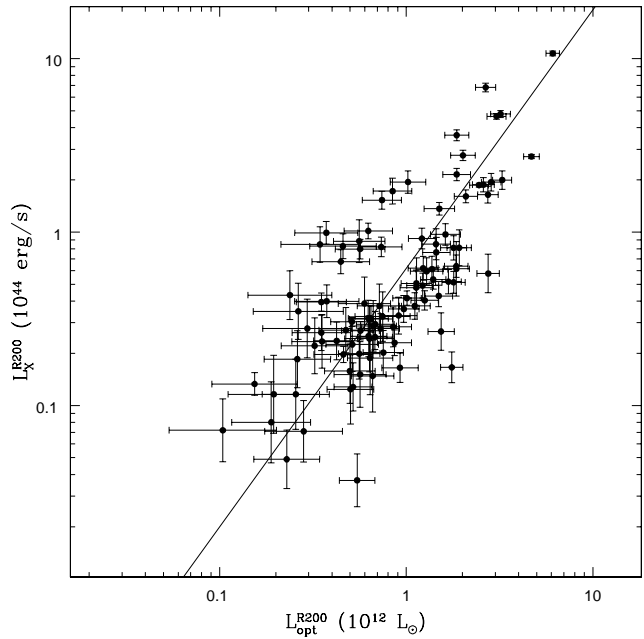}
\end{center}
\caption{Correlation between X-ray and optical luminosities, both 
measured within R200. L$_X$ is estimated with the ``annulus'' background.}
\label{fig:richscalerelations}
\end{figure}

\begin{figure}
\begin{center}
\leavevmode
\includegraphics[width=3.5in]{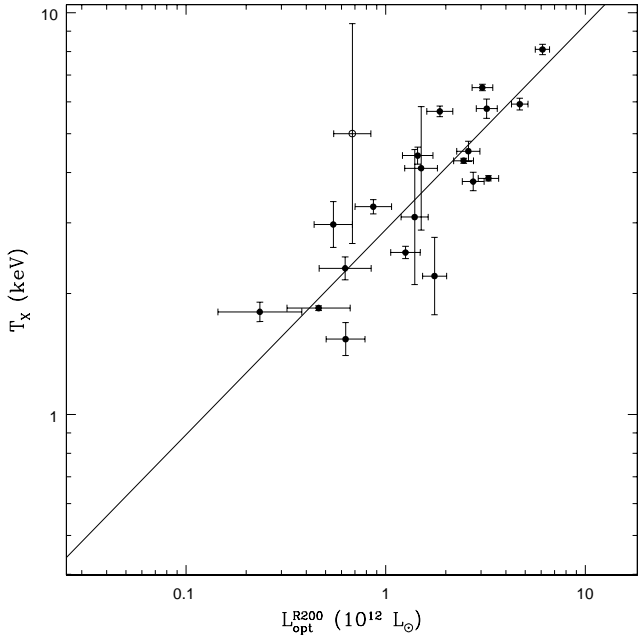}
\end{center}
\caption{Correlation between temperature and optical luminosity 
(measured within R200).}
\label{fig:richscalerelations}
\end{figure}

In this section, we investigate the correlations between optical (N$_{gals}$ 
and L$_{opt}$) and X-ray (L$_X$ and T$_X$) quantities. Note that the X-ray 
luminosity is estimated through an iterative procedure that assumes the
L$_X$-T$_X$ relation of \citet{mar98} (see paper I for more details). That
relation is close to the self-similar expectations (without any correction for 
cooling flows in L$_X$ or T$_X$). This is the same relation adopted to 
estimate L$_X$ for NORAS, 
in which T$_X \propto$ L$_X^{1/2}$. Figures 19 and 20 show the connection 
of L$_X$ to N$_{gals}$ and L$_{opt}$, respectively. In the first relation 
L$_X$ is the value obtained with the ``frame'' background, while the 
``annulus'' background is used in the latter. In both plots, 
all measurements are performed within R$_{200}$. Figure 21 shows 
the correlation 
between T$_X$ (given by BAX) and L$_{opt}^{R200}$ for 21 clusters. 

Table 9 summarizes the fitting parameters for 
the relations between X-ray and optical 
measurements within 0.50 h$^{-1}$ Mpc, R$_{200}$, and R$_{500}$. Note that
if one does not have a measure of a radius that scales with mass (R$_{500}$
or R$_{200}$), 0.50 h$^{-1}$ Mpc represents 
the optimal fixed aperture for comparing
optical and X-ray properties. \citet{pop04} and \citet{lop06} 
tested several different fixed apertures and found the scatter 
to be the smallest when using 0.50 h$^{-1}$ Mpc. The results in Table 9 are
listed for the two X-ray background types considered (``annulus'' and 
``frame''). In Table 10, we list the results involving T$_X$ (21 clusters
found in BAX) for all three apertures used for computing  N$_{gals}$ 
and L$_{opt}$. 

Our results are in good agreement with others
in the literature. For instance, POP05, also using data 
in the SDSS $r$-band, found similar results to ours: 
L$_{X,R200} \propto$ L$_{opt,R200}^{1.72}$ and 
T$_{X} \propto$ L$_{opt,R200}^{0.61}$. The results within R$_{500}$ 
and with a fixed metric (in this case for the $i$-band) are also 
close to ours. In the worst cases consistency is found within 1.5-$\sigma$. 
\citet {don01} found  similar relations but using a 
different richness definition, which hampers a direct comparison to our 
findings. 

We also find a good agreement to the T$_{X}$-N$_{gals}$ and 
T$_{X}$-L$_{opt}$ relations of \citet{lop06}. The differences are within 
1-$\sigma$. The relations involving L$_{X}$ are not as similar. 
Note that in \citet{lop06} we also force a 
richness cut, with N$_{gals} \ge 10$ (richness within 0.5 h$^{-1}$ Mpc).
\cite{lop06} also considered optical estimates 
from DPOSS and X-ray values from BAX. The heterogeneous nature of the 
luminosities derived from BAX could contribute to the discrepancies. 
However, the relations involving L$_{X}$ are much steeper and thus more 
sensitive to the sample used. If we only consider the most rich clusters
the agreement is much better than with the full sample. We also notice 
that the correlations found in the present work are in
excellent agreement with \cite {gal09}, who used DPOSS data for 
measuring N$_{gals}$ and L$_{opt}$ within 
0.50 h$^{-1}$ Mpc and directly estimated L$_{X}$ from 
RASS, as done here. 

If the density profiles for dark matter and intra-cluster gas are
self-similar, the following relations hold: M $\propto$ T$^{3/2}$, T
$\propto$ L$_X^{1/2}$ and M $\propto$ L$_X^{3/4}$. Assuming mass
traces light (constant M/L$_{opt}$) we would expect that L$_{opt}
\propto$ L$_X^{3/4}$ and L$_{opt} \propto$ T$^{3/2}$. If there is also
a strict proportionality between L$_{opt}$ and richness, the same
relations above are valid for N$_{gals}$. Our results are consistent
within 2-$\sigma$ with these relations. However, if we
assume that T $\propto$ L$_X^{2/5}$ (the typically observed result,
\citealt {dav93}) we would have L$_{opt} \propto$ L$_X^{3/5}$,
consistent with our findings at 1-$\sigma$.

\section{Connection between richness and radius}

We know that the radius and mass of a cluster scale as 
R$_{200} \propto$ M$_{200}^{1/3}$ and the number of galaxies within R 
is linked to mass as  N$_{gals,R200} \propto$ M$_{200}^\alpha$. If N$_{gals}$ 
is measured within a fixed aperture 
then N$_{gals} \propto$ N$_{gals,R200}^\beta$, with $\beta \sim 0.50-0.65$ 
(see discussion in \citealt{han05}). Hierarchical 
models of structure formation predict that the number of sub-halos within a 
system is directly proportional to the mass of the parent halo, namely 
$\alpha = $ 1.  However, different mechanisms such as a decreasing efficiency 
of star formation, or an increased merger, and destruction rate of galaxies, 
would imply a decreasing number of galaxies for higher mass halos, 
suggesting $\alpha < $ 1 (see discussion in \citealt {pop07}).  Figure 
22 shows the connection between R$_{200}$ and N$_{gals}^{R200}$, while Table 
11 has the solutions for the fits regarding the R$_{200}$-N$_{gals}^{0.5Mpc/h}$, 
R$_{200}$-N$_{gals}^{R200}$, R$_{500}$-N$_{gals}^{0.5Mpc/h}$ and
 R$_{500}$-N$_{gals}^{R500}$. 

\begin{figure}
\begin{center}
\leavevmode
\includegraphics[width=3.5in]{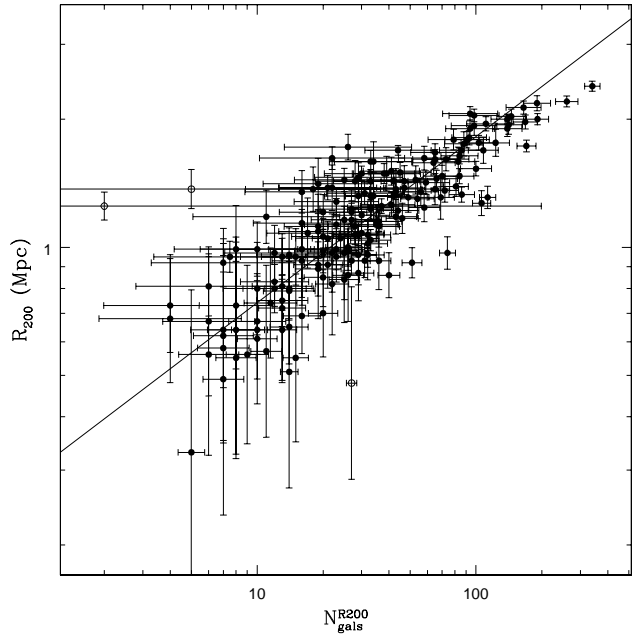}
\end{center}
\caption{Correlation between radius (R$_{200}$) and
  N$_{gals}^{R200}$.}
\label{fig:richscalerelations}
\end{figure}

We find R$_{200} \propto$ N$_{gals,R200}^{0.39}$, which indicates 
that $\alpha = 0.86$. The connection between N$_{gals}^{R200}$ and 
M$_{200}$ shown in Table 1 indicates $\alpha = 0.93$, agreeing within
1-$\sigma$ to the finding above. The connection 
between R$_{200}$ and N$_{gals}$ 
we found is also very close to the result obtained by \citet {gal09} who 
estimated R$_{200}$ photometrically (similar to \citealt{han05}), but counted 
galaxies as in the current work. This agreement corroborates
the findings and reliability of the photometric R$_{200}$ estimates given 
in \citet{gal09}. 

In Table 11 we also list the results relative to optical luminosity. We
see that the relations for L$_{opt}$ are consistent to the ones
based on richness. These results are also useful for comparison to 
the slopes of the relations from \citet{pop05}. In the SDSS $r$-band 
they found that R$_{500} \propto$ L$_{opt,R500}^{0.40}$ and 
R$_{200} \propto$ L$_{opt,R200}^{0.41}$, in good agreement
with our findings.

\section{Conclusions}

We have used a sample of 127 NoSOCS plus 56 CIRS galaxy clusters to 
investigate scaling relations at low redshift ($z \le 0.10$). For every 
cluster we previously determined (in paper I) the velocity dispersion 
($\sigma_P$), physical radii (R$_{500}$ and R$_{200}$), masses (M$_{500}$ 
and M$_{200}$), richness (N$_{gals}$), optical and X-ray luminosities 
(L$_{opt}$ and L$_X$). The last three parameters are estimated within 
0.5 h$^{-1}$ Mpc, R$_{500}$ and R$_{200}$. Substructure estimates are also 
available for nearly all clusters. We estimated the presence of substructure 
from the galaxy distribution in two and three dimensions, using the 
$\beta$ and $\Delta$ tests, respectively. For the CIRS systems we also  
consider the values of $\sigma_P$ and M$_{200}$, independently determined 
with the caustic method (RD06). For a subset of 21 clusters we have 
T$_X$ values from the literature and estimated their masses using a 
M$_{200}$-T$_X$ relation. The main conclusions we reach are:
 
\begin{enumerate}

\item Richness, optical and X-ray luminosities correlate well 
with $\sigma_p$ and mass. However, the results indicate  that the slope and 
scatter of the relations are higher when using a fixed 
aperture to compute N$_{gals}$ and L$_{opt}$. So, the most accurate 
relations are achieved when 
considering a physical radius, such as R$_{500}$ and R$_{200}$. In general,
the most accurate results are obtained within R$_{500}$. 

\item The scaling relations derived only 
with the CIRS sample are flatter than the global relations. This
indicates that the most massive systems scale differently than the poorer ones.
However, the results for the poor and rich systems are still consistent
within 1-$\sigma$.

\item The scaling relations show no significant modification when considering
only clusters without substructure. The new relations, as well as the 
scatter, are consistent with the original ones. This result
holds if we employ a two-dimensional ($\beta$) or a 
three-dimensional ($\Delta$) test. In particular, the 2D test leads to the
same scaling relations when we use the spectroscopically or photometric 
selected galaxy samples.

\item The comparison of the scaling relations obtained with optical (virial) 
and X-ray masses (from the M$_{200}$-T$_X$ relation) point to very similar 
results. In other words, the mass calibration with mass estimates from 
different wavelengths are equivalent. This is one of the main results of this 
work.

\item As the velocity dispersions and masses computed with the caustic 
technique are available in RD06 we have also derived the scaling 
relations with these parameters (considering the values of N$_{gals}$, 
L$_{opt}$ and L$_X$ computed here). Our findings are in line with the 
conclusions of paper I. Although the slope of the relations agree within 
1-$\sigma$ that is only true because the fits obtained with the caustic 
parameters have a very large scatter. For the M$_{200}$-L$_{opt}^{R200}$ 
relation the scatter in mass at fixed luminosity is (0.74 $\pm$ 0.18) when 
using the CIRS results, and only  (0.36 $\pm$ 0.06) for the virial masses
obtained in the current work. The situation is not as critical for the 
relations based on $\sigma_P$, indicating that the interloper removal 
procedure is not the 
answer for the observed discrepancies. Instead, these are due to the procedure 
used for the mass estimation, derived from the caustic mass profile. 
Uncertainties in this profile lead to different values of R$_{200}$ and
as consequence, biased results for M$_{200}$ (see discussion in paper I).

\item We find a good agreement with most of the results in the literature, even
those derived from other bands (such as 2MASS K). So, the comparison between
optical and X-ray properties, the connection between R$_{200}$ and N$_{gals}$,
and the mass calibration performed with different cluster properties, all
agree well with previous findings. This also indicates the interloper removal 
procedure we employed \citep{lop09} is robust, as other authors we compared 
to employ different techniques for selecting cluster members, or even 
determine mass from other wavelengths.

\item The scaling relations based on clusters at $z < 0.25$ are flatter than
our original results (considering only objects at $z < 0.10$). However, the
results of POP05 $-$ who use clusters in SDSS at $z < 0.25$ $-$ are
consistent with our findings. We argue that the fraction of higher-$z$ systems
in their sample is probably small, contributing little to their results. However, 
we acknowledge the relevance of this issue and conclude that clusters' 
velocity dispersion and masses must be estimated from complete spectroscopic 
samples (reaching at least M$^*+1$; see paper I). Using a large sample 
of clusters having spectra for only the bright members can severely bias 
$\sigma_P$ and mass, thus affecting the scaling relations. 

\item The main result of this paper regards the mass calibration of
galaxy clusters. We show that richness, L$_{opt}$ and L$_X$ can
reliably be used for mass estimation in the nearby
universe. Particularly, the optical properties provide slightly more
accurate relations in the present work, which is probably due to the
use of RASS in the X-ray regime (shallower than the optical data from SDSS). 
When considering all clusters, the scatter of 
M$_{500}$ at fixed N$_{gal}^{R500}$, L$_{opt}^{R500}$ and L$_{X}^{R500}$ is 
(0.43 $\pm$ 0.03), (0.44 $\pm$ 0.04) and (0.47 $\pm$ 0.05), respectively. The 
scatter of M$_{200}$ at fixed N$_{gal}^{R200}$, L$_{opt}^{R200}$ and 
L$_{X}^{R200}$ is (0.46 $\pm$ 0.04), (0.49 $\pm$ 0.04) and (0.56 $\pm$ 0.06), 
respectively. For the richer (CIRS) systems we found the scatter of 
M$_{500}$ at fixed N$_{gal}^{R500}$, L$_{opt}^{R500}$ and L$_{X}^{R500}$ is 
(0.33 $\pm$ 0.05), (0.38 $\pm$ 0.05) and (0.48 $\pm$ 0.06), respectively.
These findings
are in accord with those of POP05, who found that L$_{opt}$ is a slightly
better mass tracer than L$_X$ (also considering SDSS and RASS). This 
is a very important conclusion,
because it tells us that with accurate single band photometry we can
reliably estimate the mass of galaxy clusters, a key result
for studying the cluster mass function at low redshift. Our work also
indicates that the spectroscopic follow-up of a few dozens of clusters
to high redshifts ($z \sim 1$) can be used for the understanding of
the evolution of the scaling relations and to trace mass at high$-z$.
This is a crucial step for self-calibration methods aiming to
constrain the dark energy from the evolution of the cluster population
with cosmic time.

Note that we do not state that N$_{gals}$ or L$_{opt}$ are better mass tracers 
than L$_X$. We only say that based on RASS and SDSS that is the case. 
That is due to the fact that RASS is shallow compared to SDSS. A new 
X-ray survey, for instance based in eROSITA, would provide accurate 
X-ray luminosities for large samples of clusters. However, as we showed here, 
richness and optical luminosity are observationally cheap 
parameters that can also work as a mass proxy. This is
an important result for future large sky surveys such as 
DES, Pan-STARRS, LSST and UKIDSS. If the mass-calibration relation and
its evolution are known, these surveys can provide accurate estimates of
richness or L$_{opt}$ and thus the mass of clusters at high redshifts.

\end{enumerate}

\section*{Acknowledgments}

PAAL was supported by the Funda\c c\~ao de Amparo \`a Pesquisa do
Estado de S\~ao Paulo (FAPESP, processes 03/04110-3, 06/57027-4,
06/04955-1 and 07/04655-0). Part of this work was done at the 
Instituto Nacional de Astrof\'isica, Optica y Eletr\'onica and at the
Harvard-Smithsonian Center for Astrophysics. PAAL thanks the hospitality
during the stays in these two institutions. C. Jones thanks the Smithsonian
Astrophysical Observatory for support. The authors are thankful to 
A. Biviano for helpful discussions regarding mass estimates of 
galaxy clusters. We are thankful to Jason Pinkney for making the 
substructure codes available. We are also thankful for the valuable 
suggestions made by the referee of this paper.

This research has made use of the NASA/IPAC Extragalactic Database
(NED) which is operated by the Jet Propulsion Laboratory, California
Institute of Technology, under contract with the National Aeronautics
and Space Administration.

Funding for the SDSS and SDSS-II has been provided by the Alfred
P. Sloan Foundation, the Participating Institutions, the National
Science Foundation, the U.S. Department of Energy, the National
Aeronautics and Space Administration, the Japanese Monbukagakusho, the
Max Planck Society, and the Higher Education Funding Council for
England. The SDSS Web Site is http://www.sdss.org/.

The SDSS is managed by the Astrophysical Research Consortium for the
Participating Institutions. The Participating Institutions are the
American Museum of Natural History, Astrophysical Institute Potsdam,
University of Basel, University of Cambridge, Case Western Reserve
University, University of Chicago, Drexel University, Fermilab, the
Institute for Advanced Study, the Japan Participation Group, Johns
Hopkins University, the Joint Institute for Nuclear Astrophysics, the
Kavli Institute for Particle Astrophysics and Cosmology, the Korean
Scientist Group, the Chinese Academy of Sciences (LAMOST), Los Alamos
National Laboratory, the Max-Planck-Institute for Astronomy (MPIA),
the Max-Planck-Institute for Astrophysics (MPA), New Mexico State
University, Ohio State University, University of Pittsburgh,
University of Portsmouth, Princeton University, the United States
Naval Observatory, and the University of Washington.




\begin{table*}
\begin{minipage}{120mm}
\caption{Relations between $\sigma_P$, M$_{500}$ and 
M$_{200}$ to N$_{gals}$, using three different apertures, 0.50 h$^{-1}$ Mpc, 
R$_{500}$ and R$_{200}$. In all tables the linear fit is of the form 
${\rm ln (Y) = A + B \times ln(X/C)}$, where X and Y 
are listed in the first column. Masses are in units of 10$^{14}$ M$_{\odot}$ and
velocity dispersion in ${\rm km}$ ${\rm s^{-1}}$.
The results are also shown for three different samples, using the 127 NoSOCS 
clusters, fifty-three CIRS clusters, and the combination of the two samples 
with 180 clusters (second column). The intercept (A) and slope (B) are 
shown in the third and fourth columns, respectively. The scatter 
in the Y parameter at fixed X is in the fifth column. The total 
number of clusters for the first fit is shown as N$_{tot}$ and the number 
of clusters after a 3-$\sigma$ clip is given by N$_{use}$. The pivot point 
depends on the sample, being C = 25 for the NoSOCS and full samples and 
C = 60 for the CIRS sample.}
\label{tab:cirs56}
\begin{tabular}{llccccccc}
\hline
Relation & Sample & A & B & $\sigma_{lnY|X}$ & N$_{tot}$ & N$_{use}$ \\
X \hspace{1.2cm}  Y &  &  &  &  &  &  \\
\hline
N$_{gals}^{0.50}$ \hspace{0.70cm} $\sigma_P$ & NoSOCS & 5.88 $\pm$ 0.03 & 0.68 $\pm$ 0.04 & 0.29 $\pm$ 0.02 & 127 & 122 \vspace{0.14cm} \\
N$_{gals}^{R500}$ \hspace{0.6cm} $\sigma_P$ & NoSOCS & 5.80 $\pm$ 0.03 & 0.54 $\pm$ 0.03 & 0.23 $\pm$ 0.02 & 127 & 124 \vspace{0.14cm} \\
N$_{gals}^{R200}$ \hspace{0.6cm} $\sigma_P$ & NoSOCS & 5.70 $\pm$ 0.03 & 0.56 $\pm$ 0.03 & 0.26 $\pm$ 0.03 & 127 & 125 \vspace{0.14cm} \\
N$_{gals}^{0.50}$ \hspace{0.70cm} M$_{200}$ & NoSOCS & 0.75 $\pm$ 0.09 & 1.45 $\pm$ 0.16 & 0.69 $\pm$ 0.06 & 127 & 124 \vspace{0.14cm} \\
N$_{gals}^{R500}$ \hspace{0.6cm} M$_{500}$ & NoSOCS & 0.53 $\pm$ 0.06 & 1.01 $\pm$ 0.07 & 0.53 $\pm$ 0.05 & 127 & 126 \vspace{0.14cm} \\
N$_{gals}^{R200}$ \hspace{0.6cm} M$_{200}$ & NoSOCS & 0.36 $\pm$ 0.07 & 1.09 $\pm$ 0.09 & 0.60 $\pm$ 0.07 & 127 & 126 \vspace{0.14cm} \\
N$_{gals}^{0.50}$ \hspace{0.70cm} $\sigma_P$ & CIRS & 6.49 $\pm$ 0.03 & 0.49 $\pm$ 0.05 & 0.17 $\pm$ 0.02 & 53 & 51 \vspace{0.14cm} \\
N$_{gals}^{R500}$ \hspace{0.6cm} $\sigma_P$ & CIRS & 6.29 $\pm$ 0.02 & 0.40 $\pm$ 0.03 & 0.14 $\pm$ 0.02 & 53 & 51 \vspace{0.14cm} \\
N$_{gals}^{R200}$ \hspace{0.6cm} $\sigma_P$ & CIRS & 6.20 $\pm$ 0.03 & 0.41 $\pm$ 0.03 & 0.15 $\pm$ 0.01 & 53 & 51 \vspace{0.14cm} \\
N$_{gals}^{0.50}$ \hspace{0.70cm} M$_{200}$ & CIRS & 2.15 $\pm$ 0.08 & 1.36 $\pm$ 0.19 & 0.46 $\pm$ 0.08 & 53 & 52 \vspace{0.14cm} \\
N$_{gals}^{R500}$ \hspace{0.6cm} M$_{500}$ & CIRS & 1.54 $\pm$ 0.05 & 0.95 $\pm$ 0.09 & 0.33 $\pm$ 0.05 & 53 & 52 \vspace{0.14cm} \\
N$_{gals}^{R200}$ \hspace{0.6cm} M$_{200}$ & CIRS & 1.39 $\pm$ 0.07 & 1.00 $\pm$ 0.11 & 0.37 $\pm$ 0.05 & 53 & 52 \vspace{0.14cm} \\
N$_{gals}^{0.50}$ \hspace{0.70cm} $\sigma_P$ & ALL & 5.91 $\pm$ 0.02 & 0.69 $\pm$ 0.03 & 0.25 $\pm$ 0.02 & 180 & 171 \vspace{0.14cm} \\
N$_{gals}^{R500}$ \hspace{0.6cm} $\sigma_P$ & ALL & 5.81 $\pm$ 0.02 & 0.52 $\pm$ 0.02 & 0.19 $\pm$ 0.01 & 180 & 175 \vspace{0.14cm} \\
N$_{gals}^{R200}$ \hspace{0.6cm} $\sigma_P$ & ALL & 5.69 $\pm$ 0.02 & 0.53 $\pm$ 0.02 & 0.22 $\pm$ 0.02 & 180 & 178 \vspace{0.14cm} \\
N$_{gals}^{0.50}$ \hspace{0.70cm} M$_{200}$ & ALL & 0.79 $\pm$ 0.05 & 1.50 $\pm$ 0.08 & 0.57 $\pm$ 0.05 & 180 & 174 \vspace{0.14cm} \\
N$_{gals}^{R500}$ \hspace{0.6cm} M$_{500}$ & ALL & 0.57 $\pm$ 0.04 & 1.04 $\pm$ 0.04 & 0.43 $\pm$ 0.03 & 180 & 178 \vspace{0.14cm} \\
N$_{gals}^{R200}$ \hspace{0.6cm} M$_{200}$ & ALL & 0.37 $\pm$ 0.04 & 1.07 $\pm$ 0.04 & 0.46 $\pm$ 0.04 & 180 & 177 \vspace{0.14cm} \\

\hline
\end{tabular}
\end{minipage}
\end{table*}

\begin{table*}
\begin{minipage}{120mm}
\caption{Fit parameters for the $\sigma_P$-N$_{gals}$, M$_{200}$-N$_{gals}$ 
relations within R$_{200}$ for three more cases, all based in the NoSOCS 
sample (same pivot point as in Table 1). Masses are in units of 
10$^{14}$ M$_{\odot}$ and velocity dispersion in ${\rm km}$ ${\rm s^{-1}}$. 
Rows 1 and 2 considers
the original (instead of luminosity-weighted) coordinates. Rows 3 and 4 are 
for the case when we consider a more ``relaxed'' criteria for rejecting 
interlopers (described in $\S$ 4.1 of paper I), while rows 5 and 6 have the 
results for all the 219 NoSOCS clusters at $z \le 0.25$, instead of 
$z \le 0.10$. All the columns have the same meaning as in Table 1.}
\label{tab:cirs56}
\begin{tabular}{llccccccc}
\hline
Relation & Sample & A & B & $\sigma_{lnY|X}$ & N$_{tot}$ & N$_{use}$ \\
X \hspace{1.2cm}  Y &  &  &  &  &  &  \\
\hline
N$_{gals}^{R200}$ \hspace{0.6cm} $\sigma_P$ & NoSOCS & 5.72 $\pm$ 0.03 & 0.54 $\pm$ 0.04 & 0.29 $\pm$ 0.03 & 128 & 127 \vspace{0.14cm} \\
N$_{gals}^{R200}$ \hspace{0.6cm} M$_{200}$ & NoSOCS & 0.38 $\pm$ 0.06 & 1.05 $\pm$ 0.09 & 0.60 $\pm$ 0.07 & 128 & 127 \vspace{0.14cm} \\
N$_{gals}^{R200}$ \hspace{0.6cm} $\sigma_P$ & NoSOCS & 5.77 $\pm$ 0.03 & 0.58 $\pm$ 0.04 & 0.29 $\pm$ 0.04 & 130 & 128 \vspace{0.14cm} \\
N$_{gals}^{R200}$ \hspace{0.6cm} M$_{200}$ & NoSOCS & 0.51 $\pm$ 0.07 & 1.16 $\pm$ 0.09 & 0.66 $\pm$ 0.09 & 130 & 129 \vspace{0.14cm} \\
N$_{gals}^{R200}$ \hspace{0.6cm} $\sigma_P$ & NoSOCS & 5.63 $\pm$ 0.02 & 0.43 $\pm$ 0.02 & 0.26 $\pm$ 0.02 & 219 & 216 \vspace{0.14cm} \\
N$_{gals}^{R200}$ \hspace{0.6cm} M$_{200}$ & NoSOCS & 0.21 $\pm$ 0.04 & 0.83 $\pm$ 0.05 & 0.61 $\pm$ 0.05 & 219 & 216 \vspace{0.14cm} \\

\hline
\end{tabular}
\end{minipage}
\end{table*}

\begin{table*}
\begin{minipage}{120mm}
\caption{Analogous to Table 1, but listing the fit parameters for the 
relations of $\sigma_P$, M$_{500}$ and M$_{200}$ to L$_{opt}$. Masses are 
in units of 10$^{14}$ M$_{\odot}$ and
velocity dispersion in ${\rm km}$ ${\rm s^{-1}}$. As before,
the pivot point depends on the sample, being C = 0.40 $10^{12}$ L$_{\odot}$ 
for the NoSOCS and full samples and C = 1.10 $10^{12}$ L$_{\odot}$  for the 
CIRS sample.}
\label{tab:cirs56}
\begin{tabular}{llccccccc}
\hline
Relation & Sample & A & B & $\sigma_{lnY|X}$ & N$_{tot}$ & N$_{use}$ \\
X \hspace{1.2cm}  Y &  &  &  &  &  &  \\
\hline
L$_{opt}^{0.50}$ \hspace{0.70cm} $\sigma_P$ & NoSOCS & 5.88 $\pm$ 0.05 & 0.93 $\pm$ 0.10 & 0.41 $\pm$ 0.04 & 127 & 124 \vspace{0.14cm} \\
L$_{opt}^{R500}$ \hspace{0.6cm} $\sigma_P$ & NoSOCS & 5.79 $\pm$ 0.03 & 0.63 $\pm$ 0.04 & 0.26 $\pm$ 0.02 & 127 & 123 \vspace{0.14cm} \\
L$_{opt}^{R200}$ \hspace{0.6cm} $\sigma_P$ & NoSOCS & 5.68 $\pm$ 0.04 & 0.64 $\pm$ 0.05 & 0.38 $\pm$ 0.05 & 127 & 125 \vspace{0.14cm} \\
L$_{opt}^{0.50}$ \hspace{0.70cm} M$_{200}$ & NoSOCS & 0.76 $\pm$ 0.13 & 2.00 $\pm$ 0.32 & 0.92 $\pm$ 0.20 & 127 & 124 \vspace{0.14cm} \\
L$_{opt}^{R500}$ \hspace{0.6cm} M$_{500}$ & NoSOCS & 0.51 $\pm$ 0.06 & 1.09 $\pm$ 0.07 & 0.57 $\pm$ 0.06 & 127 & 125 \vspace{0.14cm} \\
L$_{opt}^{R200}$ \hspace{0.6cm} M$_{200}$ & NoSOCS & 0.27 $\pm$ 0.07 & 1.18 $\pm$ 0.09 & 0.63 $\pm$ 0.08 & 127 & 123 \vspace{0.14cm} \\
L$_{opt}^{0.50}$ \hspace{0.70cm} $\sigma_P$ & CIRS & 6.55 $\pm$ 0.04 & 0.59 $\pm$ 0.07 & 0.20 $\pm$ 0.03 & 53 & 52 \vspace{0.14cm} \\
L$_{opt}^{R500}$ \hspace{0.6cm} $\sigma_P$ & CIRS & 6.31 $\pm$ 0.02 & 0.41 $\pm$ 0.03 & 0.14 $\pm$ 0.01 & 53 & 50 \vspace{0.14cm} \\
L$_{opt}^{R200}$ \hspace{0.6cm} $\sigma_P$ & CIRS & 6.22 $\pm$ 0.03 & 0.39 $\pm$ 0.03 & 0.15 $\pm$ 0.02 & 53 & 51 \vspace{0.14cm} \\
L$_{opt}^{0.50}$ \hspace{0.70cm} M$_{200}$ & CIRS & 2.26 $\pm$ 0.11 & 1.48 $\pm$ 0.22 & 0.57 $\pm$ 0.10 & 53 & 53 \vspace{0.14cm} \\
L$_{opt}^{R500}$ \hspace{0.6cm} M$_{500}$ & CIRS & 1.57 $\pm$ 0.05 & 0.99 $\pm$ 0.10 & 0.38 $\pm$ 0.05 & 53 & 53 \vspace{0.14cm} \\
L$_{opt}^{R200}$ \hspace{0.6cm} M$_{200}$ & CIRS & 1.41 $\pm$ 0.06 & 0.98 $\pm$ 0.11 & 0.36 $\pm$ 0.06 & 53 & 51 \vspace{0.14cm} \\
L$_{opt}^{0.50}$ \hspace{0.70cm} $\sigma_P$ & ALL & 5.85 $\pm$ 0.03 & 0.78 $\pm$ 0.05 & 0.31 $\pm$ 0.02 & 180 & 178 \vspace{0.14cm} \\
L$_{opt}^{R500}$ \hspace{0.6cm} $\sigma_P$ & ALL & 5.76 $\pm$ 0.02 & 0.56 $\pm$ 0.02 & 0.21 $\pm$ 0.02 & 180 & 174 \vspace{0.14cm} \\
L$_{opt}^{R200}$ \hspace{0.6cm} $\sigma_P$ & ALL & 5.65 $\pm$ 0.03 & 0.55 $\pm$ 0.02 & 0.24 $\pm$ 0.02 & 180 & 176 \vspace{0.14cm} \\
L$_{opt}^{0.50}$ \hspace{0.70cm} M$_{200}$ & ALL & 0.68 $\pm$ 0.07 & 1.70 $\pm$ 0.12 & 0.72 $\pm$ 0.07 & 180 & 178 \vspace{0.14cm} \\
L$_{opt}^{R500}$ \hspace{0.6cm} M$_{500}$ & ALL & 0.49 $\pm$ 0.04 & 1.06 $\pm$ 0.04 & 0.44 $\pm$ 0.04 & 180 & 173 \vspace{0.14cm} \\
L$_{opt}^{R200}$ \hspace{0.6cm} M$_{200}$ & ALL & 0.28 $\pm$ 0.05 & 1.11 $\pm$ 0.04 & 0.49 $\pm$ 0.04 & 180 & 174 \vspace{0.14cm} \\

\hline
\end{tabular}
\end{minipage}
\end{table*}

\begin{table*}
\begin{minipage}{135mm}
\caption{Fit parameters for the $\sigma_P$-L$_X$, M$_{500}$-L$_X$, 
M$_{200}$-L$_X$ relations using two different apertures, R$_{500}$ and 
R$_{200}$. The results consider the X-ray luminosity obtained with the 
``annulus'' background (second column). Masses are in units of 
10$^{14}$ M$_{\odot}$ and
velocity dispersion in ${\rm km}$ ${\rm s^{-1}}$. First we 
show the results derived with all clusters, then only the CIRS systems 
are used. The remaining columns have the same meaning as in the previous 
tables. The pivot points are C = 0.20 $10^{44}$ erg/s for the full sampĺe and
C = 0.55 $10^{44}$ erg/s for the CIRS sampĺe.}
\label{tab:cirs56}
\begin{tabular}{lllccccccc}
\hline
Relation & Bkg Type & Sample & A & B & $\sigma_{lnY|X}$ & N$_{tot}$ & N$_{use}$ \\
X \hspace{1.2cm}  Y &  &  &  &  &  &  &  \\
\hline
L$_X^{R500}$ \hspace{0.6cm} $\sigma_P$ & Annulus & ALL & 5.90 $\pm$ 0.03 & 0.26 $\pm$ 0.03 & 0.29 $\pm$ 0.04 & 104 & 104 \vspace{0.14cm} \\
L$_X^{R200}$ \hspace{0.6cm} $\sigma_P$ & Annulus & ALL & 5.85 $\pm$ 0.04 & 0.27 $\pm$ 0.03 & 0.29 $\pm$ 0.03 & 97 & 97 \vspace{0.14cm} \\
L$_X^{R500}$ \hspace{0.6cm} M$_{500}$ & Annulus & ALL & 0.68 $\pm$ 0.06 & 0.62 $\pm$ 0.04 & 0.47 $\pm$ 0.05 & 104 & 103 \vspace{0.14cm} \\
L$_X^{R200}$ \hspace{0.6cm} M$_{200}$ & Annulus & ALL & 0.59 $\pm$ 0.08 & 0.67 $\pm$ 0.06 & 0.56 $\pm$ 0.06 & 97 & 97 \vspace{0.14cm} \\
L$_X^{R500}$ \hspace{0.6cm} $\sigma_P$ & Annulus & CIRS & 6.21 $\pm$ 0.04 & 0.18 $\pm$ 0.04 & 0.26 $\pm$ 0.03 & 52 & 52 \vspace{0.14cm} \\
L$_X^{R200}$ \hspace{0.6cm} $\sigma_P$ & Annulus & CIRS & 6.19 $\pm$ 0.04 & 0.19 $\pm$ 0.04 & 0.25 $\pm$ 0.03 & 52 & 52 \vspace{0.14cm} \\
L$_X^{R500}$ \hspace{0.6cm} M$_{500}$ & Annulus & CIRS & 1.39 $\pm$ 0.08 & 0.49 $\pm$ 0.08 & 0.48 $\pm$ 0.06 & 52 & 52 \vspace{0.14cm} \\
L$_X^{R200}$ \hspace{0.6cm} M$_{200}$ & Annulus & CIRS & 1.40 $\pm$ 0.08 & 0.52 $\pm$ 0.08 & 0.49 $\pm$ 0.06 & 52 & 52 \vspace{0.14cm} \\

\hline
\end{tabular}
\end{minipage}
\end{table*}

\begin{table*}
\begin{minipage}{135mm}
\caption{Analogous to the previous table, but for the results based in 
the ``frame'' background. Masses are in units of 10$^{14}$ M$_{\odot}$ and
velocity dispersion in ${\rm km}$ ${\rm s^{-1}}$.}
\label{tab:cirs56}
\begin{tabular}{lllccccccc}
\hline
Relation & Bkg Type & Sample & A & B & $\sigma_{lnY|X}$ & N$_{tot}$ & N$_{use}$ \\
X \hspace{1.2cm}  Y &  &  &  &  &  &  &  \\
\hline
L$_X^{R500}$ \hspace{0.6cm} $\sigma_P$ & Frame & ALL & 5.92 $\pm$ 0.03 & 0.24 $\pm$ 0.02 & 0.28 $\pm$ 0.04 & 98 & 98 \vspace{0.14cm} \\
L$_X^{R200}$ \hspace{0.6cm} $\sigma_P$ & Frame & ALL & 5.90 $\pm$ 0.04 & 0.22 $\pm$ 0.03 & 0.28 $\pm$ 0.03 & 90 & 90 \vspace{0.14cm} \\
L$_X^{R500}$ \hspace{0.6cm} M$_{500}$ & Frame & ALL & 0.73 $\pm$ 0.06 & 0.59 $\pm$ 0.04 & 0.46 $\pm$ 0.04 & 98 & 96 \vspace{0.14cm} \\
L$_X^{R200}$ \hspace{0.6cm} M$_{200}$ & Frame & ALL & 0.68 $\pm$ 0.08 & 0.61 $\pm$ 0.06 & 0.55 $\pm$ 0.05 & 90 & 89 \vspace{0.14cm} \\
L$_X^{R500}$ \hspace{0.6cm} $\sigma_P$ & Frame & CIRS & 6.20 $\pm$ 0.04 & 0.18 $\pm$ 0.04 & 0.26 $\pm$ 0.03 & 50 & 50 \vspace{0.14cm} \\
L$_X^{R200}$ \hspace{0.6cm} $\sigma_P$ & Frame & CIRS & 6.19 $\pm$ 0.04 & 0.18 $\pm$ 0.03 & 0.25 $\pm$ 0.03 & 49 & 49 \vspace{0.14cm} \\
L$_X^{R500}$ \hspace{0.6cm} M$_{500}$ & Frame & CIRS & 1.38 $\pm$ 0.08 & 0.49 $\pm$ 0.07 & 0.49 $\pm$ 0.06 & 50 & 50 \vspace{0.14cm} \\
L$_X^{R200}$ \hspace{0.6cm} M$_{200}$ & Frame & CIRS & 1.40 $\pm$ 0.08 & 0.50 $\pm$ 0.07 & 0.50 $\pm$ 0.06 & 49 & 49 \vspace{0.14cm} \\

\hline
\end{tabular}
\end{minipage}
\end{table*}

\begin{table*}
\begin{minipage}{140mm}
\caption{Fit parameters for the M$_{200}$-N$_{gals}$ and M$_{200}$-L$_{opt}$ for 
clusters with or without substructure. Masses are in units of 
10$^{14}$ M$_{\odot}$ and
velocity dispersion in ${\rm km}$ ${\rm s^{-1}}$. 
First, we show (lines 1-6) the results 
considering an aperture of R$_{200}$ for all clusters and for the substructure 
free systems. These are selected with the $\Delta$ or $\beta$ tests, applied 
only to galaxies considered as cluster members ($\S$4). Then we show the 
results considering the same aperture and the 2-D test ($\beta$), but using 
the photometric sample (no restriction to galaxies with spectra). That is 
seen in rows 7-10. Finally, we list the results with a fixed metric of 
1.5 h$^{-1}$ Mpc for estimating substructure and using the $\beta$ test. 
Again, the photometric data are considered (rows 11-14). The parameters 
of each relation are listed in column 1, while the substructure test (when 
it is the case) is shown in column 2. The remaining columns have the same 
meaning as in the previous tables. The pivot points for all samples are 
C = 25 when considering richness, and C = 0.40 $10^{12}$ L$_{\odot}$ 
for L$_{opt}$.}
\label{tab:cirs56}
\begin{tabular}{lccccccccc}
\hline
Relation & Sub-test & Sample & A & B & $\sigma_{lnY|X}$ & N$_{tot}$ & N$_{use}$ \\
X \hspace{1.2cm}  Y &  &  &  &  &  &  &  \\
\hline
N$_{gals}^{R200}$ \hspace{0.6cm} M$_{200}$ &  & ALL & 0.40 $\pm$ 0.05 & 1.10 $\pm$ 0.05 & 0.48 $\pm$ 0.04 & 170 & 167 \vspace{0.14cm} \\
N$_{gals}^{R200}$ \hspace{0.6cm} M$_{200}$ & $\Delta$ & Sub-free & 0.40 $\pm$ 0.05 & 1.10 $\pm$ 0.06 & 0.49 $\pm$ 0.05 & 131 & 129 \vspace{0.14cm} \\
N$_{gals}^{R200}$ \hspace{0.6cm} M$_{200}$ & $\beta$ & Sub-free & 0.40 $\pm$ 0.05 & 1.10 $\pm$ 0.05 & 0.49 $\pm$ 0.04 & 144 & 141 \vspace{0.14cm} \\
L$_{opt}^{R200}$ \hspace{0.65cm} M$_{200}$ &  & ALL & 0.31 $\pm$ 0.05 & 1.12 $\pm$ 0.05 & 0.51 $\pm$ 0.04 & 170 & 163 \vspace{0.14cm} \\
L$_{opt}^{R200}$ \hspace{0.65cm} M$_{200}$ & $\Delta$ & Sub-free & 0.35 $\pm$ 0.06 & 1.08 $\pm$ 0.07 & 0.52 $\pm$ 0.06 & 131 & 127 \vspace{0.14cm} \\
L$_{opt}^{R200}$ \hspace{0.65cm} M$_{200}$ & $\beta$ & Sub-free & 0.33 $\pm$ 0.05 & 1.12 $\pm$ 0.05 & 0.53 $\pm$ 0.04 & 144 & 138 \vspace{0.14cm} \\

\hline
N$_{gals}^{R200}$ \hspace{0.6cm} M$_{200}$ &  & ALL & 0.38 $\pm$ 0.04 & 1.09 $\pm$ 0.04 & 0.48 $\pm$ 0.03 & 179 & 176 \vspace{0.14cm} \\
N$_{gals}^{R200}$ \hspace{0.6cm} M$_{200}$ & $\beta$ & Sub-free & 0.38 $\pm$ 0.05 & 1.06 $\pm$ 0.05 & 0.48 $\pm$ 0.04 & 136 & 133 \vspace{0.14cm} \\
L$_{opt}^{R200}$ \hspace{0.65cm} M$_{200}$ &  & ALL & 0.28 $\pm$ 0.05 & 1.12 $\pm$ 0.05 & 0.52 $\pm$ 0.04 & 179 & 173 \vspace{0.14cm} \\
L$_{opt}^{R200}$ \hspace{0.65cm} M$_{200}$ & $\beta$ & Sub-free & 0.33 $\pm$ 0.06 & 1.07 $\pm$ 0.06 & 0.51 $\pm$ 0.05 & 136 & 131 \vspace{0.14cm} \\

\hline
N$_{gals}^{0.50}$ \hspace{0.70cm} M$_{200}$ &  & ALL & 0.80 $\pm$ 0.05 & 1.54 $\pm$ 0.09 & 0.64 $\pm$ 0.06 & 183 & 177 \vspace{0.14cm} \\
N$_{gals}^{0.50}$ \hspace{0.70cm} M$_{200}$ & $\beta$ & Sub-free & 0.79 $\pm$ 0.06 & 1.66 $\pm$ 0.12 & 0.59 $\pm$ 0.07 & 119 & 117 \vspace{0.14cm} \\
L$_{opt}^{0.50}$ \hspace{0.76cm} M$_{200}$ &  & ALL & 0.69 $\pm$ 0.07 & 1.74 $\pm$ 0.13 & 0.79 $\pm$ 0.08 & 183 & 181 \vspace{0.14cm} \\
L$_{opt}^{0.50}$ \hspace{0.76cm} M$_{200}$ & $\beta$ & Sub-free & 0.70 $\pm$ 0.07 & 1.72 $\pm$ 0.13 & 0.70 $\pm$ 0.08 & 119 & 116 \vspace{0.14cm} \\

\hline
\end{tabular}
\end{minipage}
\end{table*}

\begin{table*}
\begin{minipage}{120mm}
\caption{Fit parameters for the relations connecting M$_{200}$ and $\sigma_P$
with N$_{gals}$, L$_{opt}$ and L$_X$. These last parameters are determined 
within R$_{200}$. Masses are in units of 10$^{14}$ M$_{\odot}$ and
velocity dispersion in ${\rm km}$ ${\rm s^{-1}}$. 
The parameters of each relation are listed in column 1. 
In the second column, we indicate the method used for estimating mass or 
$\sigma_P$. Mass can be derived from the M$_{200}$-T$_X$ relation of 
POP05 or from the caustic (RD06). Velocity dispersion values come
from the latter (RD06). Relations derived with the first 
method (masses estimated from 
T$_X$) are based in the 21 clusters with temperature available. Results
obtained with the second method (considering the caustic values) use the
53 CIRS clusters. The pivot points are the same as listed in the previous 
tables when considering the CIRS sample. For the 21 rich clusters with
T$_X$ available the pivot points are assumed the same. The remaining columns
have the same meaning as in the tables above.}
\label{tab:cirs56}
\begin{tabular}{lcccccccc}
\hline
Relation & Mass & A & B & $\sigma_{lnY|X}$ & N$_{tot}$ & N$_{use}$ \\
X \hspace{1.2cm}  Y &  &  &  &  &  &  \\
\hline
N$_{gals}^{R200}$ \hspace{0.55cm} M$_{200}$ & X-ray & 0.90 $\pm$ 0.10 & 0.92 $\pm$ 0.10 & 0.34 $\pm$ 0.08 & 21 & 20 \vspace{0.14cm} \\
L$_{opt}^{R200}$ \hspace{0.6cm} M$_{200}$ & X-ray & 0.99 $\pm$ 0.10 & 0.85 $\pm$ 0.11 & 0.33 $\pm$ 0.08 & 21 & 20 \vspace{0.14cm} \\
L$_X^{R200}a$ \hspace{0.44cm} M$_{200}$ & X-ray & 1.15 $\pm$ 0.14 & 0.39 $\pm$ 0.10 & 0.36 $\pm$ 0.11 & 20 & 20 \vspace{0.14cm} \\
L$_X^{R200}b$ \hspace{0.47cm} M$_{200}$ & X-ray & 1.02 $\pm$ 0.12 & 0.47 $\pm$ 0.08 & 0.25 $\pm$ 0.07 & 19 & 18 \vspace{0.14cm} \\

\hline
N$_{gals}^{R200}$ \hspace{0.55cm} $\sigma_P$ & Caustic & 6.34 $\pm$ 0.03 & 0.27 $\pm$ 0.06 & 0.18 $\pm$ 0.02 & 53 & 52 \vspace{0.14cm} \\
L$_{opt}^{R200}$ \hspace{0.57cm} $\sigma_P$ & Caustic & 6.36 $\pm$ 0.03 & 0.25 $\pm$ 0.05 & 0.18 $\pm$ 0.03 & 53 & 52 \vspace{0.14cm} \\
L$_X^{R200}a$ \hspace{0.42cm} $\sigma_P$ & Caustic & 6.36 $\pm$ 0.03 & 0.12 $\pm$ 0.03 & 0.21 $\pm$ 0.03 & 52 & 51 \vspace{0.14cm} \\
L$_X^{R200}b$ \hspace{0.47cm} $\sigma_P$ & Caustic & 6.35 $\pm$ 0.03 & 0.12 $\pm$ 0.03 & 0.21 $\pm$ 0.03 & 49 & 48 \vspace{0.14cm} \\
N$_{gals}^{R200}$ \hspace{0.53cm} M$_{200}$ & Caustic & 0.64 $\pm$ 0.14 & 1.35 $\pm$ 0.34 & 0.77 $\pm$ 0.22 & 53 & 53 \vspace{0.14cm} \\
L$_{opt}^{R200}$ \hspace{0.57cm} M$_{200}$ & Caustic & 0.72 $\pm$ 0.12 & 1.26 $\pm$ 0.29 & 0.74 $\pm$ 0.18 & 53 & 53 \vspace{0.14cm} \\
L$_X^{R200}a$ \hspace{0.42cm} M$_{200}$ & Caustic & 0.73 $\pm$ 0.13 & 0.56 $\pm$ 0.19 & 0.87 $\pm$ 0.12 & 52 & 52 \vspace{0.14cm} \\
L$_X^{R200}b$ \hspace{0.46cm} M$_{200}$ & Caustic & 0.68 $\pm$ 0.12 & 0.54 $\pm$ 0.15 & 0.87 $\pm$ 0.12 & 49 & 49 \vspace{0.14cm} \\

\hline
\end{tabular}
{$a-$ L$_X^{R200}$ estimated with the ``annulus'' background;
$b-$ L$_X^{R200}$ estimated with the ``frame'' background}
\end{minipage}
\end{table*}

\begin{table*}
\begin{minipage}{85mm}
\caption{Slope and orthogonal scatter of different scaling relations, from 
this work or from other authors. Masses are in units of 
10$^{14}$ M$_{\odot}$ and velocity dispersion in ${\rm km}$ ${\rm s^{-1}}$. 
The parameters involved in each relation 
are listed in column 1; the sample considered is in column 2; the source in 
columns 3; while the slope and scatter are in columns 4 and 5, respectively.}
\label{tab:cirs56}
\begin{tabular}{lcccc}
\hline
Relation & Sample & Source & B & $\sigma_{lnY|X}$ \\
X \hspace{1.2cm}  Y &  &  &  &  \\
\hline
N$_{gals}^{R500}$ \hspace{0.55cm} M$_{500}$ & ALL & This work & 1.04 & 0.43 \vspace{0.14cm} \\
N$_{gals}^{R200}$ \hspace{0.55cm} M$_{200}$ & ALL & This work & 1.07 & 0.46 \vspace{0.14cm} \\
L$_{opt}^{R500}$ \hspace{0.6cm} M$_{500}$ & ALL & This work & 1.06  & 0.44 \vspace{0.14cm} \\
L$_{opt}^{R200}$ \hspace{0.6cm} M$_{200}$ & ALL & This work & 1.11  & 0.49 \vspace{0.14cm} \\
L$_{opt}^{R500}$ \hspace{0.6cm} M$_{500}$ & Extended & Pop05 & 1.09  & 0.50 \vspace{0.14cm} \\
L$_{opt}^{R200}$ \hspace{0.6cm} M$_{200}$ & Extended & Pop05 & 1.06  & 0.51 \vspace{0.14cm} \\

\hline
L$_X^{R500}a$ \hspace{0.45cm} M$_{500}$ & ALL & This work & 0.62  & 0.47 \vspace{0.14cm} \\
L$_X^{R200}a$ \hspace{0.45cm} M$_{200}$ & ALL & This work & 0.67  & 0.56 \vspace{0.14cm} \\
L$_X^{R500}b$ \hspace{0.5cm} M$_{500}$ & ALL & This work & 0.59  & 0.46 \vspace{0.14cm} \\
L$_X^{R200}b$ \hspace{0.5cm} M$_{200}$ & ALL & This work & 0.61  & 0.55 \vspace{0.14cm} \\
L$_X$ \hspace{0.97cm} M$_{500}$ & Extended & Pop05 & 0.59  & 0.59 \vspace{0.14cm} \\
L$_X$ \hspace{0.97cm} M$_{200}$ & Extended & Pop05 & 0.60  & 0.54 \vspace{0.14cm} \\
L$_X$ \hspace{0.97cm} M & --- & Sta06 & 0.63  & 0.43 \vspace{0.14cm} \\
L$_X$ \hspace{0.97cm} M$_{500}$ & --- & Vik08 & 0.62  & 0.25 \vspace{0.14cm} \\
L$_X$ \hspace{0.97cm} M & --- & Ett04 & 0.53  & 0.25 \vspace{0.14cm} \\

\hline
L$_{opt}^{R500}$ \hspace{0.6cm} $\sigma_P$ & ALL & This work & 0.55  & 0.24 \vspace{0.14cm} \\
L$_{opt}^{R200}$ \hspace{0.6cm} $\sigma_P$ & ALL & This work & 0.56  & 0.21 \vspace{0.14cm} \\
L$_{opt}^{R500}$ \hspace{0.6cm} $\sigma_P$ & Extended & Pop05 & 0.43  & 0.17 \vspace{0.14cm} \\
L$_{opt}^{R200}$ \hspace{0.6cm} $\sigma_P$ & Extended & Pop05 & 0.43  & 0.17 \vspace{0.14cm} \\

\hline
L$_X^{R500}a$ \hspace{0.45cm} $\sigma_P$ & ALL & This work & 0.26  & 0.29 \vspace{0.14cm} \\
L$_X^{R500}b$ \hspace{0.5cm} $\sigma_P$ & ALL & This work & 0.24  & 0.28 \vspace{0.14cm} \\
L$_X$ \hspace{0.97cm} $\sigma_P$ & Extended & Pop05 & 0.27  & 0.25 \vspace{0.14cm} \\

\hline
\end{tabular}
{$a-$ L$_X^{R200}$ estimated with the ``annulus'' background;
$b-$ L$_X^{R200}$ estimated with the ``frame'' background. The works of 
Pop05, Sta06, Vik08 and Ett04 are \citet{pop05, sta06, vik08} and 
\citet{ett04}, respectively. Note that the results from \citet{sta06} are 
dependent of the cosmology. When considering results from the third year of 
WMAP, they find a reduced scatter $\sigma_{lnM|L_{X}} = 0.25$, consistent 
to \citet{vik08}.}
\end{minipage}
\end{table*}

\begin{table*}
\begin{minipage}{135mm}
\caption{Fit parameters for the relations between X-ray luminosity (L$_X$) 
and optical properties (N$_{gals}$ and L$_{opt}$) within three different 
apertures. The aperture used for computing L$_X$ is always the same as 
indicated for N$_{gals}$ or L$_{opt}$. L$_X$ is measured in units of 
$10^{44}$ erg/s. The background type considered 
for the L$_X$ estimates is listed in the second column. The sample used for
deriving the relations is shown
in the third column. The remaining columns are as in Table 1. The pivot 
points are C = 25 when considering richness, and 
C = 0.40 $10^{12}$ L$_{\odot}$ for L$_{opt}$.}
\label{tab:cirs56}
\begin{tabular}{lllccccccc}
\hline
Relation & Bkg Type & Sample & A & B & $\sigma_{lnY|X}$ N$_{tot}$ & N$_{use}$ \\
X \hspace{1.2cm}  Y &  &  &  &  &  &  &  \\
\hline
N$_{gals}^{0.50}$ \hspace{0.70cm} L$_X$ & Annulus  & ALL & -1.51 $\pm$ 0.09 & 2.28 $\pm$ 0.20 & 0.83 $\pm$ 0.13 & 105 & 105 \vspace{0.14cm} \\
N$_{gals}^{R500}$ \hspace{0.6cm} L$_X$ & Annulus  & ALL & -1.69 $\pm$ 0.09 & 1.64 $\pm$ 0.10 & 0.62 $\pm$ 0.09 & 104 & 103 \vspace{0.14cm} \\
N$_{gals}^{R200}$ \hspace{0.6cm} L$_X$ & Annulus  & ALL & -1.85 $\pm$ 0.12 & 1.59 $\pm$ 0.12 & 0.63 $\pm$ 0.09 & 97 & 97 \vspace{0.14cm} \\
L$_{opt}^{0.50}$ \hspace{0.76cm} L$_X$ & Annulus  & ALL & -1.68 $\pm$ 0.12 & 2.38 $\pm$ 0.23 & 0.97 $\pm$ 0.19 & 105 & 104 \vspace{0.14cm} \\
L$_{opt}^{R500}$ \hspace{0.66cm} L$_X$ & Annulus  & ALL & -1.77 $\pm$ 0.10 & 1.62 $\pm$ 0.11 & 0.65 $\pm$ 0.10 & 104 & 103 \vspace{0.14cm} \\
L$_{opt}^{R200}$ \hspace{0.66cm} L$_X$ & Annulus  & ALL & -1.84 $\pm$ 0.13 & 1.49 $\pm$ 0.12 & 0.64 $\pm$ 0.10 & 97 & 97 \vspace{0.14cm} \\

\hline
N$_{gals}^{0.50}$ \hspace{0.70cm} L$_X$ & Frame  & ALL & -1.47 $\pm$ 0.09 & 2.24 $\pm$ 0.20 & 0.82 $\pm$ 0.13 & 103 & 103 \vspace{0.14cm} \\
N$_{gals}^{R500}$ \hspace{0.6cm} L$_X$ & Frame  & ALL & -1.71 $\pm$ 0.09 & 1.69 $\pm$ 0.11 & 0.68 $\pm$ 0.10 & 98 & 98 \vspace{0.14cm} \\
N$_{gals}^{R200}$ \hspace{0.6cm} L$_X$ & Frame  & ALL & -1.88 $\pm$ 0.12 & 1.62 $\pm$ 0.12 & 0.75 $\pm$ 0.11 & 90 & 90 \vspace{0.14cm} \\
L$_{opt}^{0.50}$ \hspace{0.76cm} L$_X$ & Frame  & ALL & -1.63 $\pm$ 0.12 & 2.34 $\pm$ 0.23 & 0.96 $\pm$ 0.20 & 103 & 102 \vspace{0.14cm} \\
L$_{opt}^{R500}$ \hspace{0.66cm} L$_X$ & Frame  & ALL & -1.73 $\pm$ 0.10 & 1.61 $\pm$ 0.11 & 0.68 $\pm$ 0.11 & 98 & 97 \vspace{0.14cm} \\
L$_{opt}^{R200}$ \hspace{0.66cm} L$_X$ & Frame  & ALL & -1.99 $\pm$ 0.13 & 1.59 $\pm$ 0.13 & 0.78 $\pm$ 0.14 & 90 & 89 \vspace{0.14cm} \\

\hline
\end{tabular}
\end{minipage}
\end{table*}

\begin{table*}
\begin{minipage}{115mm}
\caption{Fit parameters for the relations between X-ray temperature T$_X$ 
(in keV) and optical properties (N$_{gals}$ and L$_{opt}$) 
within three different apertures. Only the 21 clusters with x-ray 
temperature available in BAX are used in the fits. That is indicated in
the second column (``sample''). The remaining columns are as in the 
previous table. The pivot points are C = 60 when considering richness, and 
C = 1.10 $10^{12}$ L$_{\odot}$ for L$_{opt}$.}
\label{tab:cirs56}
\begin{tabular}{llccccccc}
\hline
Relation & Sample & A & B & $\sigma_{lnY|X}$ & N$_{tot}$ & N$_{use}$ \\
X \hspace{1.2cm}  Y &  &  &  &  &  &  \\
\hline
N$_{gals}^{0.50}$ \hspace{0.70cm} T$_X$ & BAX & 1.52 $\pm$ 0.05 & 0.73 $\pm$ 0.11 & 0.20 $\pm$ 0.05 & 21 & 21 \vspace{0.14cm} \\
N$_{gals}^{R500}$ \hspace{0.6cm} T$_X$ & BAX & 1.22 $\pm$ 0.06 & 0.50 $\pm$ 0.07 & 0.19 $\pm$ 0.04 & 21 & 21 \vspace{0.14cm} \\
N$_{gals}^{R200}$ \hspace{0.6cm} T$_X$ & BAX & 1.05 $\pm$ 0.06 & 0.56 $\pm$ 0.06 & 0.21 $\pm$ 0.05 & 21 & 20 \vspace{0.14cm} \\
L$_{opt}^{0.50}$ \hspace{0.76cm} T$_X$ & BAX & 1.56 $\pm$ 0.07 & 0.69 $\pm$ 0.13 & 0.23 $\pm$ 0.05 & 21 & 21 \vspace{0.14cm} \\
L$_{opt}^{R500}$ \hspace{0.66cm} T$_X$ & BAX & 1.21 $\pm$ 0.06 & 0.52 $\pm$ 0.07 & 0.20 $\pm$ 0.04 & 21 & 20 \vspace{0.14cm} \\
L$_{opt}^{R200}$ \hspace{0.66cm} T$_X$ & BAX & 1.10 $\pm$ 0.06 & 0.51 $\pm$ 0.06 & 0.21 $\pm$ 0.04 & 21 & 20 \vspace{0.14cm} \\

\hline
\end{tabular}
\end{minipage}
\end{table*}

\begin{table*}
\begin{minipage}{120mm}
\caption{Fit parameters for the  R$_{500}$-N$_{gals}$, R$_{200}$-N$_{gals}$, 
R$_{500}$-L$_{opt}$ and R$_{200}$-L$_{opt}$ relations using three 
different apertures for computing N$_{gals}$ or L$_{opt}$. The physical radii
are measured in Mpc. The remaining columns are as in the previous table. The 
pivot points are C = 25 when considering richness, and 
C = 0.40 $10^{12}$ L$_{\odot}$ for L$_{opt}$.}
\label{tab:cirs56}
\begin{tabular}{llccccccc}
\hline
Relation & Sample & A & B & $\sigma_{lnY|X}$ & N$_{tot}$ & N$_{use}$ \\
X \hspace{1.2cm}  Y &  &  &  &  &  &  \\
\hline
N$_{gals}^{0.50}$ \hspace{0.70cm} R$_{500}$ & ALL & -0.12 $\pm$ 0.02 & 0.47 $\pm$ 0.02 & 0.19 $\pm$ 0.01 & 180 & 173 \vspace{0.14cm} \\
N$_{gals}^{R500}$ \hspace{0.60cm} R$_{500}$ & ALL & -0.19 $\pm$ 0.01 & 0.37 $\pm$ 0.01 & 0.14 $\pm$ 0.01 & 180 & 175 \vspace{0.14cm} \\
N$_{gals}^{0.50}$ \hspace{0.70cm} R$_{200}$ & ALL & 0.20 $\pm$ 0.02 & 0.48 $\pm$ 0.02 & 0.19 $\pm$ 0.01 & 180 & 173 \vspace{0.14cm} \\
N$_{gals}^{R200}$ \hspace{0.60cm} R$_{200}$ & ALL & 0.05 $\pm$ 0.02 & 0.39 $\pm$ 0.02 & 0.17 $\pm$ 0.01 & 180 & 177 \vspace{0.14cm} \\

\hline
L$_{opt}^{0.50}$ \hspace{0.70cm} R$_{500}$ & ALL & -0.16 $\pm$ 0.02 & 0.56 $\pm$ 0.04 & 0.24 $\pm$ 0.02 & 180 & 178 \vspace{0.14cm} \\
L$_{opt}^{R500}$ \hspace{0.60cm} R$_{500}$ & ALL & -0.22 $\pm$ 0.01 & 0.40 $\pm$ 0.02 & 0.17 $\pm$ 0.01 & 180 & 175 \vspace{0.14cm} \\
L$_{opt}^{0.50}$ \hspace{0.70cm} R$_{200}$ & ALL & 0.16 $\pm$ 0.02 & 0.56 $\pm$ 0.04 & 0.24 $\pm$ 0.02 & 180 & 178 \vspace{0.14cm} \\
L$_{opt}^{R200}$ \hspace{0.60cm} R$_{200}$ & ALL & 0.01 $\pm$ 0.02 & 0.40 $\pm$ 0.02 & 0.18 $\pm$ 0.01 & 180 & 176 \vspace{0.14cm} \\

\hline
\end{tabular}
\end{minipage}
\end{table*}



\appendix

\section[]{Update values of X-ray luminosities}

We fixed a bug in the code used to estimate X-ray luminosity and its 
associated error. Hence we update these values in the table A1 below, which 
replaces Table 5 from paper I. The meaning of all columns is as in paper I.
In the first column we give the cluster name; then in the next three columns
we list the values of L$_X$ (and its associated error) for the three 
backgrounds (``annulus'', ``frame'' and ``boxes'', respectively). In the 
fifth column we show the X-ray temperature measure from BAX (when available),
while the interpolated temperature obtained from the L$_X$-T$_X$ relation
is shown in the sixth column. The last column indicates whether we had a 
significant detection (SD), an upper limit (UL), or if the measurements 
should be taken with concern (XX). The cluster is marked as XX if it is too
close to a border, or if the background error or net count rates could not
be determined. The last two columns of the table are obtained with the 
``annulus'' background.

\begin{table*}
\begin{minipage}{160mm}
\caption{X-ray luminosity and temperature of the 183 NoSOCS plus CIRS clusters.}
\label{tab:cirs56}
\begin{tabular}{@{}lccccccc}
\hline
 name & L$_X$ (annulus) &  L$_X$ (frame) &  L$_X$ (box) & T$_X$ (BAX) & T$_X$ (interp. RASS) & NOTE \\
 & (10$^{44}$erg s$^{-1}$) & (10$^{44}$erg s$^{-1}$) & (10$^{44}$erg s$^{-1}$) & (keV) & (keV) & \\
\hline
NSCS J121847+484410 & 0.049 $\pm$  0.019 & 0.052 $\pm$  0.020 & 0.051 $\pm$  0.019 & ------ & 0.7 & SD \\
NSCS J011502+002441 & 0.404 $\pm$  0.052 & 0.497 $\pm$  0.058 & 0.497 $\pm$  0.056 & 2.53 $_{-0.09} ^{+0.09}$ & 2.1 & SD \\
NSCS J100242+324218 & 0.395 $\pm$  0.048 & 0.342 $\pm$  0.045 & 0.371 $\pm$  0.046 & ------ & 2.1 & SD \\

\hline
\end{tabular}
{Note. - A portion of this table is shown here for guidance regarding its form and content. A full version is available in the electronic edition of the MNRAS.}
\end{minipage}
\end{table*}

\label{lastpage}

\end{document}